\documentclass[11pt]{article}
\providecommand{\keywords}[1]{\textbf{\textit{Keywords:---}} #1}

\usepackage{authblk}
\usepackage{amssymb}
\usepackage{amsmath,amsthm}
 \usepackage{graphicx}
\usepackage{color}

\def\si {{\sigma}}

\def\ga{{ \gamma}}

\def\eps{{ \epsilon}}

\newcommand{\nn}{\nonumber}
\newcommand{\dis}{\displaystyle}

\newtheorem{thm}{Theorem}

\newtheorem{lem}{\indent Lemma}

\newcommand{\mmmintone}[1]{{\dis{\int\kern -.36cm-}}_{\kern-.21cm\substack{#1}}\;\;}
\newcommand{\mmmintwo}[2]{{\dis{\int\kern -.43cm-}}_{\kern-.21cm\substack{#1}}^{\substack{#2}}\;\;}
\newcommand{\submint}{{\scriptstyle{\int\kern -.66em -}}}
\newcommand{\submintone}[1]{{\scriptstyle{\int\kern -.66em-}}_{\scriptscriptstyle{\kern-.21em\substack{#1}}}}
\newcommand{\fracmint}{{\textstyle{\int\kern -.88em -}}}
\newcommand{\fracmintone}[1]{{\textstyle{\int\kern -.88em
-}}_{\scriptscriptstyle{\kern-.21em\substack{#1}}}\;}

\title{Particle models with self sustained current}

\author{M. Colangeli \footnote{Universit\`{a} degli Studi dell'Aquila, Via Vetoio, 67100 L'Aquila, Italy.\\ E-mail: matteo.colangeli1@univaq.it}, A. De Masi\footnote{Universit\`{a} degli Studi dell'Aquila, Via Vetoio, 67100 L'Aquila, Italy.\\ E-mail: anna.demasi@univaq.it}, E. Presutti \footnote{Gran Sasso Science Institute, Viale F. Crispi 7, 00167 L' Aquila, Italy.\\ E-mail: errico.presutti@gmail.com}}

\date{\today}

\begin{document}

\maketitle

\begin{abstract}
\noindent
We present some computer simulations run on a stochastic CA (cellular automaton).  The CA simulates a gas of particles which are in a channel,
the interval $[1,L]$ in $\mathbb Z$, but   also in ``reservoirs'' $\mathcal R_1$ and $\mathcal R_2$. The evolution in the channel simulates a lattice gas with Kawasaki dynamics with attractive Kac interactions; the temperature is chosen smaller than the mean field critical one.  There are also exchanges of particles between the channel and the reservoirs and among reservoirs.  When the rate of exchanges among reservoirs is in a suitable interval the CA reaches an apparently stationary state with a non zero current; for different choices of the initial condition the current changes sign.  We have a quite satisfactory theory of the phenomenon but we miss a full mathematical proof.
\end{abstract}

\noindent

\section{Introduction}
\label{sec.0}

In this paper we  introduce
models of macroscopic dissipative systems
made of interacting particles which move stochastically in a circuit and exhibit a very surprising behavior.
Despite the fact that there is no external bias
  we see, after a transient, an apparently stationary state with a
non zero current, with  suitably different initial conditions we may select another state  with the opposite value
of the current.
%
%
%
%
%
We speculate that on much longer times there is
a ``dynamical phase transition'' with the   two states alternating one after the other.  To make an analogy with
equilibrium phase transitions, consider
the 2D Ising model in a large but finite box with nearest neighbor ferromagnetic interactions. Running the Glauber dynamics at a temperature below the critical value we  typically see long time intervals  where the magnetization density has approximately the plus equilibrium value alternating via tunneling with those where it is close to the minus equilibrium
value.  The analogue of the equilibrium magnetization in our model is the current as we have two states with opposite values of the current. However we observe  our circuit for times long but much smaller than those
for tunneling so that we only see one of the two currents (selected by the initial condition) which then
looks stationary.
Our analysis
relies mostly on computer simulations, we have theoretical explanations but we
miss a mathematical proof.

There is a huge literature on the more general question of existence of periodic motions
or oscillations especially in the context of biological systems and chemical reactions, the classical reference is the book by Kuramoto, \cite{kuramoto}.
We just quote here a few examples selected with the purpose of introducing what we will be doing in this paper.

 In \cite{tass} P. Tass discusses a simple system of rotators which interact attractively with each other and are subject to white noise forces.  For small interactions the stationary state is homogeneous and even though each particle rotates there is no macroscopic change.  However if the interaction increases the rotators form a macroscopic
cluster which then moves periodically.  This is a simplified model for
neural activities, the angle of the rotator is related to the neuron potential and the crossing from $2\pi$ to 0 is interpreted as the neuron  discharging
its potential (``firing''): the appearance of a cluster causes a great potential change when the cluster
crosses $2\pi$ which could explain some diseases related
to anomalous neuron
firing.

A quantum analogue of the rotator model has been studied by Wilczek in \cite{wilczek1} where it is shown that there are
ground states with a localized cluster  which rotates, this phenomenon called a
``time crystal''.  Comments on time crystals can be found in \cite{bruno}.
Experimental evidence of
``time crystals''   are presented in \cite{zhang}. Time crystals in a classical (i.e.\ non quantum) context have been considered in \cite{wilczek2}.

A rotators model is also considered in \cite{giacomin} where  an additional
external force is present.  The main point in the paper is to show that for a critical set of values of the
parameters there is a cluster which is however blocked (by the external force).  However if the white noise strength is increased then the cluster starts moving and performs a periodic motion,  this being a nice example  of noise-induced periodicity. Also in our models noise is the fuel which makes the system run.

\begin{figure}[h]
\centering
\includegraphics[width=0.5 \textwidth]{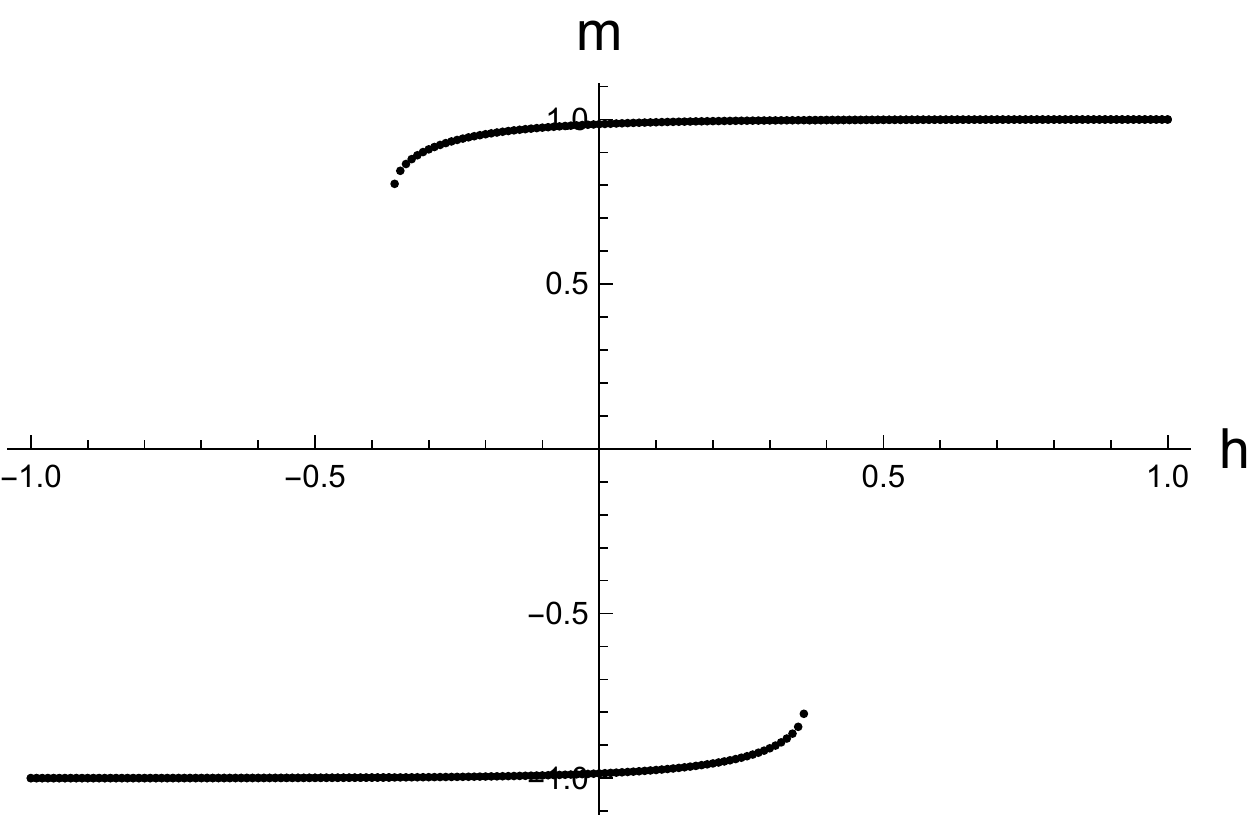}
\caption{{\footnotesize Hysteresis cycle, with $\beta=2.5$.}}
\label{fig:1}
\end{figure}

In the above models each particle by itself rotates: the macroscopic rotations
arise from a ``phase synchronization'' of the rotators. Instead in the FitzHugh Nagumo class of models
for the firing cycles of a neuron,
the appearance of periodic motions is due
to a  different, more intrinsic mechanism.  For what follows it is convenient to consider a particular model in the
class which can and will be read in a statistical mechanics language.  In such a context the model is defined by two (macroscopic) variables, the magnetization $m$ and the magnetic field $h$. $m$ is the ``fast'' and $h$ the  ``slow variable'' as the  evolution is defined by the equations:
\begin{equation}
\label{0.1}
\eps \frac{dm}{dt} = -m + \tanh\{\beta (m + h)\},\quad  \frac{dh}{dt} =-m
\end{equation}
where $\eps>0$ is the ``small parameter'' and $\beta>1$ the inverse temperature.  It can be seen that \eqref{0.1} has
a (stable)
periodic solution  which in the limit $\eps\to 0$ becomes
the hysteresis cycle: $m= m_\pm(h)$, $m_{+}(h)$ the positive solution of $m = \tanh\{\beta (m + h)\}$ which exists for $h>-h_c$, $h_c>0$;  $m_{-}(h)=-m_{+}(-h)$, $h<h_c$, see Fig.  \ref{fig:1}.  The transition from the upper curve $m_{+}(\cdot)$ to the lower one $m_{-}(\cdot)$ (and viceversa) is discontinuous and hence very sharp
for $\eps>0$ small, a fact which catches the main feature of the neuron  voltage cycle namely that at the firing the potential changes very abruptly.
Observe
that $m_+(h)$ is metastable for $h<0$ as well as
$m_-(h)$   for $h>0$, the metastable values of the magnetization will play a fundamental role also in this paper.

Dai Pra et al., \cite{daipra}, derived similar patterns in a macroscopic limit
from a  Ising spin model with  mean field interactions giving nice examples of ``intrinsic'' periodic oscillations in the stochastic Ising model.  In this paper we will consider the relaxed version of mean field as defined by Kac potentials.


All the above examples can be interpreted in terms of a current in a circuit but in all of them there is a more or less hidden bias because the current can flow only in one direction and not in the opposite one, so that they do not fit in what we are looking for.  However they have all a
common feature with our models, namely the presence of a phase transition, responsible in the rotator models for the formation of a cluster and in the FitzHugh Nagumo models for the presence of a hysteresis cycle.  The way phase transitions appear in our analysis is the following.  In a first order phase transition there is a spontaneous separation of phases which gives rise to gradients of the order parameter without currents being present.  The Fourier law associates to a gradient a current (in the opposite direction) so that the phase transition generates  ``effective
forces'' which prevent the gradients to give rise to currents. Our idea is to exploit such forces to construct a ``battery'' which allows for a non zero current in a circuit.

Our battery is a cellular automaton which simulates the Kawasaki dynamics in a lattice gas
with interactions given by  an attractive Kac potential which in the Lebowitz-Penrose limit has a van der Waals phase transition. Therefore we can distinguish between stable, metastable and  unstable values of the density.
The main and
somehow unexpected feature of the system is that if we connect the endpoints of the channel to ``infinite'' (i.e.\ true) reservoirs which fix
the  density at values $\rho_- $ and $\rho_+= 1-\rho_->\rho_-$ with $\rho_{\pm}$ metastable densities  we observe numerically a  current which goes through the channel from the reservoir with smaller density $\rho_-$ to the one with the larger density $\rho_+$.  We have a theoretical explanation of the phenomenon in terms of properties of the solution of an integral equation obtained from the  process in the ``mesoscopic limit'' where the scaling parameter $\ga$ of the Kac potential vanishes, but we could verify these properties only numerically.


In \cite{CDP} we have presented numerical evidence that the current in the CA flows from the reservoir with smaller density to the one with larger density.  In this paper we present a more complete set of simulations from where a very complex structure
emerges for which we have a theoretical explanation,
but we miss
a complete mathematical proof. The other main point in this paper is
that we can exploit the above to
construct a circuit with a self sustained current  without  an external bias, as claimed in the first sentence of this Introduction.  This is obtained by
making the reservoirs finite and allowing also particles exchanges among the reservoirs.
We show (via the simulations) that for suitable values of the parameters
there are initial conditions which give rise to a steady
non zero current (stationary for the times of our simulations); there are also other initial conditions where the current flows in the opposite direction and
still others where there is no current at all.  The  state with zero current seems unstable
while those with a non zero current seem locally stable.

As suggested by a referee, similar phenomena have also been studied in other models, e.g. the Bunimovich's mushroom billiard model \cite{Bunim01} in which the presence of peculiar transport regimes can be traced back to the lack of ergodicity of the microscopic dynamics;
but we have not yet explored this issue.\\

The paper is organized as follows. In Section \ref{sec.j1} we define two different versions of the CA used in the simulations, namely: one describing a single (open) channel in contact with two reservoirs (hereafter called OS-CA), and another mimicking the particle dynamics in a closed circuit (called CC-CA). \\
In Section \ref{sec.j1.4} we present the results of the simulations obtained by running the OS-CA and also explain how to run the CC-CA by exploiting the results first obtained with the OS-CA.\\
In Section \ref{sec.j1.1} we illustrate the behavior of the particle current in the CC-CA and comment on the dependence of this quantity on the parameters of the model. \\
In Section \ref{sec.j1.3} we study the continuum (mesoscopic) limits of both the OS-CA and the CC-CA, which are described by an integro-differential equation; proofs are deferred to the Appendix \ref{A1}. \\
In Section \ref{sec.j1.5} we discuss the adiabatic limit of the model, and check the consistency of our simulations of the CC-CA with the predicted adiabatic behavior.\\
In Section \ref{sec6} we consider the case where the reservoirs have stable densities and in Section \ref{sec7}  where the densities are not stable.\\
In Section \ref{sec.j9} we study the stability of a stationary density profile, referred to below as \textit{the ``bump'' solution}, close to the boundary. \\
Concluding remarks are finally drawn in Section \ref{sec:concl}.

\setcounter{equation}{0}

\section{The cellular automata}
		\label{sec.j1}

In this section we define two cellular automata: the first one,  called ``open system cellular automaton'', OS-CA in short, has been first
introduced in
\cite{LOP} and then  used in \cite{CDP} to simulate a system in contact with reservoirs.  The second one, simply
called ``closed circuit  cellular automaton'', CC-CA,  is a modification of the first one obtained by making finite the reservoirs and adding direct exchanges
between them, so that it simulates a closed circuit.

\medskip

\noindent{\bf The OS-CA.} The  OS-CA describes the evolution of  particles in a ``channel" $ \{1,2,..,L\}$,  $L>1$ a positive integer.  Besides moving in the channel particles may also leave from or enter into the channel through $L$ and $1$ (we then say that they are absorbed or released from  the reservoir  $\mathcal R_2$ if this happens at $L$ and  from reservoir $\mathcal R_1$ if it happens at $1$).
The two reservoirs are ``infinite" in the sense that they do not have memory of the particles which are absorbed or released.

The CA in the channel is a parallel updating version of a weakly asymmetric
simple exclusion process,
designed for computer simulations. The $d=1$ symmetric simple exclusion process is a system of random walks
jumping to the right and left with equal probability, the jump being suppressed
if the arrival site is occupied.  The weak asymmetry that we add is a
small bias to jump in the direction where the density is higher. If the channel was   a torus this would produce  a phase separation into a region where the density is higher and another where it is smaller.  But our channel  is open as particles may leave or enter into the channel in a setup typical of the Fourier law but in a context where phase transitions are present.

Let us now go back to the definition of the CA.
The phase space is $\mathcal S=\{(x,v), x\in\{1,..,L\}, v\in\{-1,1\}\}$, particles configurations are functions $\eta:\mathcal S\to \{0,1\}$, $\eta(x,v) \in \{0,1\}$ denotes the occupation variable at $(x,v)$ and $v$ will be interpreted as a velocity. $\eta(x)=\eta(x,-1)+\eta(x,1)\in\{0,1,2\}$ denotes the total number of particles at $x$. We may add a suffix $t$ when the occupation variables are computed at time $t$.

The definition of the OS-CA involves four more parameters: $\ga^{-1}\in \mathbb N$,
$C>0$ and $\rho_{\pm}\in [0,1]$. In the simulations presented in this paper we have fixed
$\ga^{-1}=30$, $C=1.25$, while the length of the channel is set equal to $L=600$. $\rho_{\pm}$ are referred to as the density of reservoir  $\mathcal R_2$, respectively  $\mathcal R_1$, they are fixed during a simulation but they may be changed in different simulations. In the definition of the CA we will use the notation
\begin{equation}
\label{j1.2}
N_{+,x,\ga}
= \sum_{y=x+1}^{ x+\ga^{-1}}\eta^{(+)}(y),\;
N_{-,x,\ga}
= \sum_{y= x-\ga^{-1}}^{x-1}\eta^{(-)}(y),\quad x \in [1,L]
\end{equation}
where $\eta^{(+)}(y)= \eta(y)$ if $y \in [1,L]$
and $\eta^{(+)}(y)= 2\rho_{+}$ if $y >L$; similarly  $\eta^{(-)}(y)= \eta(y)$ if $y \in [1,L]$
and $\eta^{(-)}(y)= 2\rho_{-}$ if $y <1$. We want $N_{+,x,\ga}$ to be  the total number of particles to the right of $x$ within distance $\ga^{-1}$ from $x$, however it may happen that if $x$ is close to the right boundary then there are not  $\ga^{-1}$ sites in the channel to the right of $x$. Suppose that there are only  $\ga^{-1}-m$ such sites, we then add fictitiously $2m$ phase points $(y,v)$, $v=\pm 1$ and $y$ takes  $m$ values to be thought as $m$ physical sites to the right of the channel.
The occupation number $\eta(y,v)$ is then set equal to   $\rho_+$
so that the contribution to $N_{+,x,\ga}$ of the extra $m$ sites is $2\rho_+ m$, which explains the factor 2 in the definition of $\eta^{(+)}(y)$. Analogous interpretation applies to $\eta^{(-)}(y)$.

We are now ready to define how the OS-CA operates:
the unit time step updating (from $t$ to $t+1$) is obtained as the result of three successive
operations, we denote by $\eta$ the configuration at time $t$, by $\eta'$ and $\eta''$ two consecutive updates starting from $\eta$
and by $\eta'''$ the final update which gives the configuration at time $t+1$.

\begin{enumerate}
\item {\em velocity flip}.  At all sites $x\in \{1,..,L\}$ where there is only
one particle we update its velocity  to become $+1$ with probability $\frac 12
+ \eps_{x,\ga}$ and $-1$ with probability $\frac 12
- \eps_{x,\ga}$, $\eps_{x,\ga}= C\ga^2[N_{+,x,\ga}-N_{-,x,\ga}]$ (the definition is well posed because $(2\ga^{-1}) C\ga^2= 2.5/30<\frac 12$, $(2\ga^{-1}) $ being an upper bound for $|N_{+,x,\ga}-N_{-,x,\ga}|$).
At all other sites the occupation numbers are left unchanged. We denote by $\eta'$ the occupation numbers after the flip.

\item {\em advection}. After deleting the particles in the channel at $(1,-1)$ and $(L,1)$
(if present) we  let
each one of the remaining particles in the channel move by one lattice step in the direction of its velocity.
  We denote by $\eta''$ the occupation numbers after this advection step.

\item {\em exchanges with the reservoirs}.  With probability $\rho_+$ we put a particle at $(L,-1)$ and with probability  $1-\rho_+$ we leave  $(L,-1)$ empty.  We do independently the same operations at $(1,1)$ but with $\rho_-$ instead of $\rho_+$.  The final configuration is then denoted by  $\eta'''$.

\end{enumerate}

\noindent{\bf The CC-CA.}  We now turn to the second CA which
describes the evolution of  particles in a ``closed circuit". The phase space is the disjoint union   
$\mathcal S  \cup \mathcal R_1\cup \mathcal R_2$, where $\mathcal S$ is as before 
while the two reservoirs
$\mathcal R_1$ and $\mathcal R_2$ are
finite sets both with cardinality $R$, $R$ a positive, even integer. $R$ is interpreted as the number of phase points in the reservoir, thus there will be $R/2$ sites  with velocity $1$ and  $R/2$ sites
with velocity $-1$: the velocities in the reservoirs however do not play any role in the evolution, they are used only to have a symmetric description of the channel and the reservoirs.  Unlike in the OS-CA now the total number of particles (i.e.\ those in the channel and in the reservoirs) is constant in time.  In the CC-CA the densities $\rho_{\pm}$ in the two reservoirs are no longer constant but given by $N_{\mathcal R_1}/R$ and $N_{\mathcal R_2}/R$ where
	\begin{equation}
		\label{j1.1}
N_{\mathcal R_1} =  \sum_{(x,v)\in \mathcal R_1}\eta(x,v),\quad
N_{\mathcal R_2} =   \sum_{(x,v)\in \mathcal R_2}\eta(x,v)
	\end{equation}
Accordingly we define
$N_{\pm,x,\ga}$ in the CA  as in \eqref{j1.2} but with $\rho_{\pm}$  replaced by the instantaneous values
$N_{\mathcal R_1}/R$ and $N_{\mathcal R_2}/R$ of the density in $\mathcal R_1$ and $\mathcal R_2$. With these notation  the first two steps of the evolution in the CA are the same as in the OS-CA. We call again $\eta'$ and $\eta''$ the configurations in the system after the first and the second step, with $\eta''=\eta'=\eta$ in $\mathcal R_1 \cup \mathcal R_2$ (i.e.\ the occupation numbers in the reservoirs are unchanged in the first two steps).  In the  third step   instead they may change as we are going to see.

\medskip
\noindent
{\em 3. The new third step, (reservoirs exchanges).} Its definition involves  a new, suitably small parameter $\ga p > 0$.
We  first select with uniform probability a phase point $(x_1,v_1)\in \mathcal R_1$ and   $(x_2,v_2)\in \mathcal R_2$: if $\eta(x_1,v_1)=0$ we set $\eta'''(1,1)=0$, if instead $\eta(x_1,v_1)=1$ we set $\eta'''(1,1)=1$.  Analogously $\eta'''(L,-1)=0,1$ if  $\eta(x_2,v_2)=0,1$.  This concludes the definition of $\eta'''$ in the channel while in the reservoirs $\eta''' = \theta'''$, with $\theta'''$ defined as follows.  We first define $\theta'$ by setting $\theta'(x,v)=\eta(x,v)$ for $(x,v)$ in $\mathcal R_1$ with $(x,v)  \ne (x_1,v_1)$ and $\theta'(x_1,v_1)=0$.  $\theta'(x,v)$ is defined analogously in  $\mathcal R_2$. $\theta''(x,v)$ is obtained from $\theta'(x,v)$ by adding a particle in the first empty point of $\mathcal R_1$ (according to a fixed but arbitrary order) if $\eta'(1,-1)=1$, otherwise $\theta''=\theta'$ in  $\mathcal R_1$.
$\theta''$ is defined analogously in  $\mathcal R_2$.  Finally $\theta'''$ is obtained from $\theta''$ in the following way.  With probability $1-\ga p $ we let $\theta'''=\theta''$
while with  probability $\ga p $ we do the following: we choose with uniform probability
$(y_1,v_1)\in \mathcal R_1$ and $(y_2,v_2)\in \mathcal R_2$ and exchange $\theta''(y_1,v_1)$ with $\theta''(y_2,v_2)$. To be well defined we have tacitly supposed that $ \ga p  \le 1$, actually $ \ga p  \ll 1$ in the simulations.

Heuristically $\ga p$ is the rate at which particles jump directly from  a reservoir to the other. Without the channel these exchanges would eventually make the  densities of the two reservoirs equal to each other.

\medskip

\noindent{\bf Magnetization variables.}

To exploit the symmetries in the system it is
convenient to introduce spin variables. We set in the CC-CA:
\begin{equation}
\label{j1.13}
  \si(x) = \eta(x,1)+\eta(x,-1)-1
\end{equation}
both  in the channel and in the reservoirs.
(possibly adding $t$ when the variables are computed at time $t$).  We call $\dis{S_{\rm ch}= \sum_{x=1}^L\si(x)}$ the total spin in the channel, thus
	\begin{equation}\label{2.4aA}
S_{\rm ch} = N_{\rm ch}-L,\qquad N_{\rm ch}:=\sum_{x=1}^L\eta(x)
	\end{equation}
Recalling \eqref{j1.1}, we define  analogously to \eqref{2.4aA}
		\begin{equation}
	\label{j1.20.3.1}
 S_{\mathcal R_1}= N_{\mathcal R_1} - \frac R2,\quad S_{\mathcal R_2}=N_{\mathcal R_2}- \frac R2
		\end{equation}
We define also the  magnetization density in the two reservoirs
	\begin{equation}
	\label{2.5aa}
m^{CC}_{-} =\frac { S_{\mathcal R_1}}{R/2}
,\qquad 	m^{CC}_{+} =\frac { S_{\mathcal R_2}}{R/2}
	\end{equation}
 In the OS-CA the magnetization density in the ``reservoirs''  is
 \begin{equation}
	\label{2.5bb}m_{\pm}= 2\rho_{\pm} -1\end{equation}

\medskip

\noindent{\bf Currents.} For the OS-CA we define $j_{\mathcal R_1\to {\rm ch}}(t)$ and $j_{{\rm ch} \to\mathcal R_2}(t)$ as the number of particles which go from
$\mathcal R_1$ to the channel  minus those which go
from the channel to $\mathcal R_1$ in the time step $t\to t+1$ and respectively,  the number of particles which go from the channel
to $\mathcal R_2$  minus those which go
from  $\mathcal R_2$ to the channel in the time step $t\to t+1$. Thus
 \begin{eqnarray}
&&j_{\mathcal R_1\to {\rm ch}}(t)= \eta'''(1,1;t) - \eta'(1,-1;t)\nn
\\ \label{2.7aa} \\&&\nn
j_{{\rm ch} \to\mathcal R_2}(t)= \eta'(L,1;t)-\eta'''(L,-1;t)
\end{eqnarray}
with $\eta'$,  $\eta''$ and  $\eta'''$ the occupation numbers after the three updates
which lead from $t$ to $t+1$.

In the CC-CA the currents $j^{CC}_{\mathcal R_1\to {\rm ch}}(t)$ and $j^{CC}_{{\rm ch} \to\mathcal R_2}(t) $ are defined  by the same expression as in \eqref{2.7aa} with the  new  $\eta$'s. The current between the reservoirs is defined  as the number of particles which go from
$\mathcal R_2$ to $\mathcal R_1$ minus those which go
from $\mathcal R_1$ to $\mathcal R_2$ in the time step $t\to t+1$, thus:
	\begin{eqnarray}
	\label{2.9ab}
 j_{\mathcal R_2\to \mathcal R_1}(t)
= -\sum_{(x,v)\in \mathcal R_2}  [\theta'''(x,v;t)-\theta''(x,v;t)]
\end{eqnarray}

\bigskip

\noindent{\bf  Conservation laws.} In the OS-CA we have
	\begin{equation}\label{2.9aA}
 N_{\rm ch}(t+1)- N_{\rm ch}(t)= j_{\mathcal R_1\to {\rm ch}}(t)-j_{{\rm ch} \to\mathcal R_2}(t)
	\end{equation}
In the CC-CA the analogue of \eqref{2.9aA} holds as well:
	\begin{equation}\label{2.10aA}
N_{\mathcal R_1} (t+1)- N_{\mathcal R_1} (t)=  j^{CC}_{\mathcal R_2\to \mathcal R_1}(t)-
j^{CC}_{\mathcal R_1\to {\rm ch}}(t)	\end{equation}
with analogous formula for $N_{\mathcal R_2}$. As a consequence, in the CC-CA, the total number of particles  $N_{\mathcal R_1} +N_{\mathcal R_1} + N_{\rm ch}$ is conserved as well as the total spin $S_{\mathcal R_1} +S_{\mathcal R_1} + S_{\rm ch}$.

\bigskip

\noindent{\bf  Initial conditions. } In
the OS-CA we impose $\rho_+> \rho_-$, $\rho_++\rho_-=1$. Observe that this implies $m_+>0$ and $m_++m_-=0$.
Analogously in the CC-CA  we initially impose that
\begin{equation}
\label{j1.3.1}
N_{\mathcal R_1}  + N_{\mathcal R_2}  = R, \quad N_{\mathcal R_2} > \frac R2
\end{equation}
The initial state in the channel will be specified in the sequel.
\medskip

\noindent{\bf Parameters of the simulations.} We conclude the section by recalling the values of the parameters that will be used in the simulations:
\begin{equation}
\label{j1.3.0}
\ga^{-1}=30, \;C=1.25, \;\beta=2.5,\;L=600,\;
R= 10^5,\quad R=:\ga^{-1}a,\;L =: \ga^{-1} \ell
\end{equation}

\vskip1cm

\setcounter{equation}{0}

\section {The  OS-CA}
\label{sec.j1.4}

In this section we present the simulations obtained by
running  the OS-CA,
recall  that this CA has been defined in terms of two fixed densities $\rho_+$ and $\rho_-$, $\rho_++\rho_-=1$,  which in the magnetization variables, see \eqref{2.5bb}, amounts to fix  $m_+>0$, $m_-=-m_+$.
As already mentioned the  OS-CA simulates the typical Fourier law experiments therefore the physically most  relevant quantity is the stationary current $j(m_+)$: $j=j(m_+)$
plays the role of the equation of state in a non equilibrium context (due to the presence of the reservoirs) and defines the ``non equilibrium thermodynamics'' of the system.

Thus our first task is to consider the currents in the CA, since the
  instantaneous currents defined in \eqref{2.7aa} are strongly fluctuating, we  take averages:
\begin{eqnarray}
\label{j1.20.2.1}
&&j^{T}_{\mathcal R_1\to {\rm ch}}= \frac 1{T}\sum_{t=0}^{T-1}
j_{ \mathcal R_1\to {\rm ch}}(t)
\end{eqnarray}
 In general with $f^T$ we will denote the average of $f$, thus $j^{T}_{{\rm ch} \to\mathcal R_2}$ is the averaged current from the channel to $\mathcal R_2$.

Strictly speaking stationarity is reached as $T\to \infty$, existence of the limit should follow (almost everywhere) from the Birkhoff theorem.  Of course in the simulations we cannot take such a limit and the value of $T$ is chosen empirically in such a way that $j^{T}_{{\rm ch} \to\mathcal R_2}
\approx j^{T}_{\mathcal R_1\to {\rm ch}}$
looks independent of $T$. The initial condition in the channel is with all phase points empty, we have checked that with other conditions the final current does not change appreciably.  The stationarity condition  $j^{T}_{{\rm ch} \to\mathcal R_2}(m_+)
- j^{T}_{\mathcal R_1\to {\rm ch}}(m_+)\approx 0$ is also satisfied, typical values are $10^{-8}$ while the  currents have order $10^{-5}$. $10^{-8}$ is also considerably smaller than the a-priori bound
\begin{equation*}
|j^{T}_{\mathcal R_1\to {\rm ch}}- j^T_{ {\rm ch}\to \mathcal R_2}| \le \frac{2L}{T} = 1.2 \cdot 10^{-7}, \text{when $T= 10^{10}$}
\end{equation*}

\begin{figure}[h!]
\centering
\includegraphics[width=0.70 \textwidth]{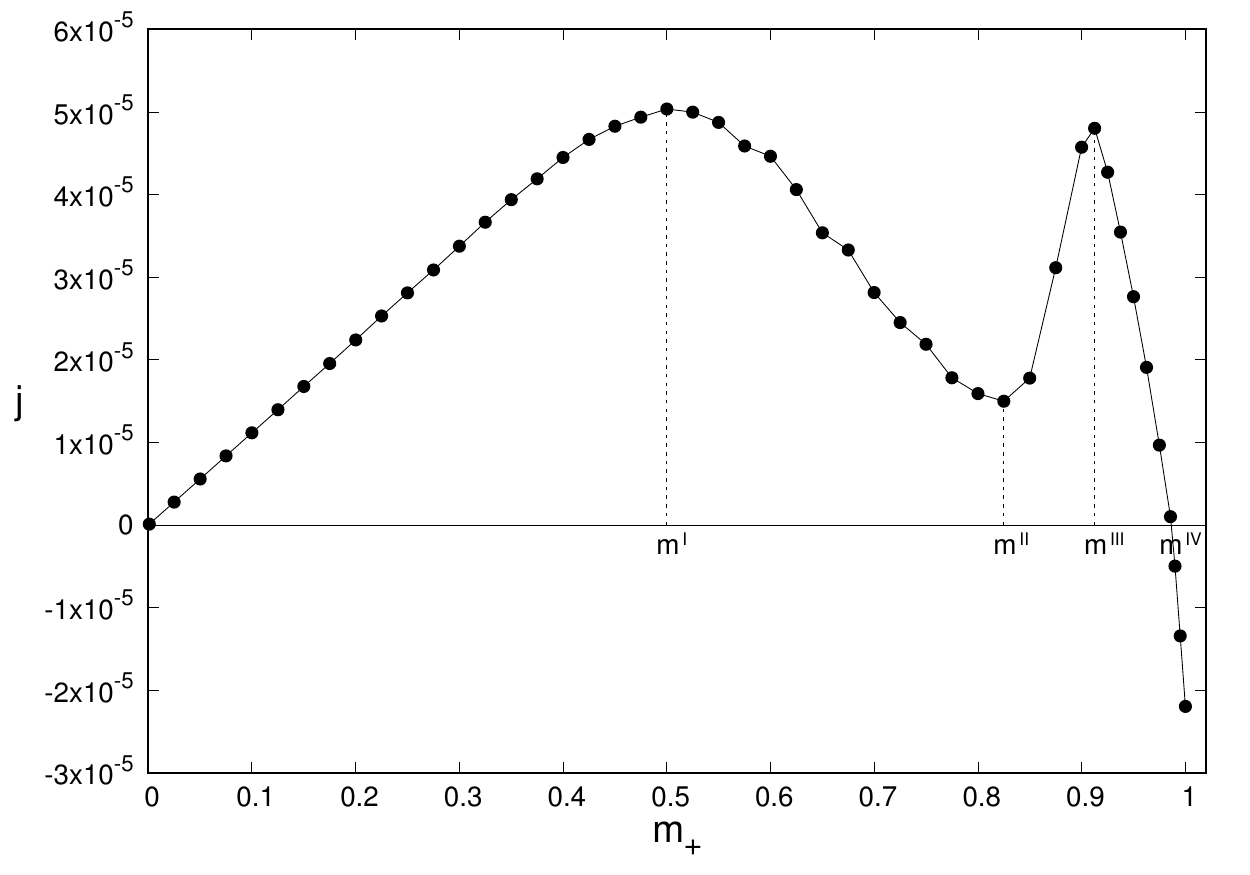}
\caption{{\footnotesize
We plot $j:=j^{T}_{{\rm ch} \to\mathcal R_2}$ as a function of $m_+$ (black dots). The continuous line is $j(m_+)$. Shown are the values $m'= 0.500$, $m''=0.825$, $m'''=0.912$, $m^{iv}= 0.985$.}}
\label{fig:2}
\end{figure}

\begin{figure}[h!]
\centering
\includegraphics[width=0.70 \textwidth]{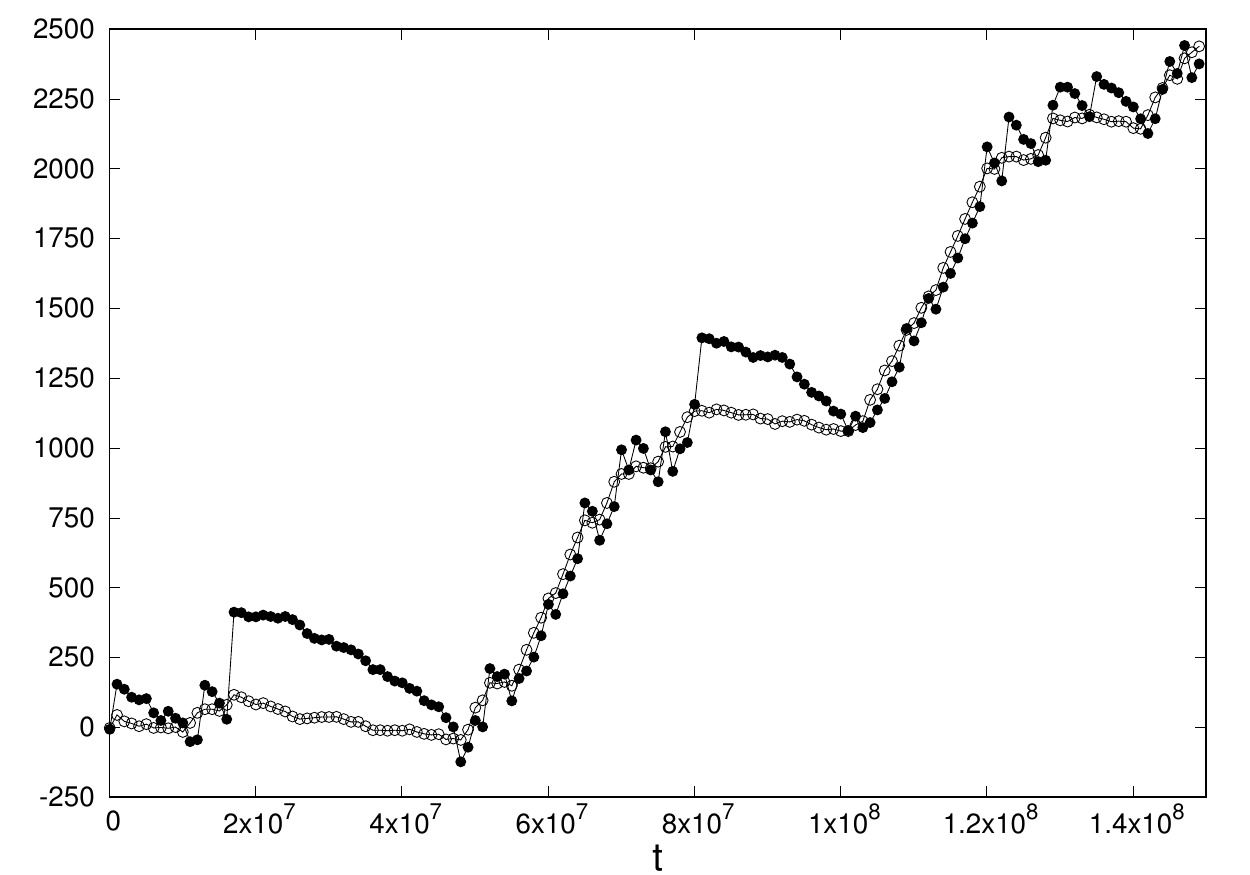}
\caption{{\footnotesize
We plot $j^{t}_{\mathcal R_1\to {\rm ch}}(m_+)t$ (black circles) and $j^t_{ {\rm ch}\to \mathcal R_2}(m_+)t$ (empty circles) as functions of time $t$ , with $m_+ \in [m',m''']$.}}
\label{fig:3}
\end{figure}

The  black dots in Fig. \ref{fig:2} are the  values of $j^{T}_{{\rm ch} \to\mathcal R_2}(m_+)$
in the simulations  done with $T= 3 \cdot 10^9$ for $m_+ \in (0,m')$ and $m_+>m'''$ and $T= 10^{10}$ elsewhere.
The continuous line in Fig. \ref{fig:2}, denoted by $j(m_+)$, is a continuous interpolation of $j^{T}_{{\rm ch} \to\mathcal R_2}(m_+)$ which we presume to be a good approximation of simulations done with the other values of $m_+$, it is therefore the ``experimental'' value for the non equilibrium equation of state $j=j(m_+)$.
The
main features  in   Fig.  \ref{fig:2}  (where $m'= 0.500$, $m''= 0.825$, $m'''=0.912$  and $m^{iv}= 0.985$)) are:

  \begin{itemize}

  \item  For $m_+\in (m^{iv},1]$ the current  $j(m_+)$ is negative  in agreement with the Fourier law, while for $m_+<m^{iv}$ the current is positive going from  smaller to larger values of the magnetization (i.e. from $m_-$ to $m_+$).

   \item  $j(m_+)$ is first increasing till $m'$, then   decreasing till $m''$, again increasing till $m'''$ and finally decreasing then after.

 \item In Fig. \ref{fig:3} we plot $j^{t}_{\mathcal R_1\to {\rm ch}}(m_+)t$ and $j^t_{ {\rm ch}\to \mathcal R_2}(m_+)t$, $t \le T$ with $m_+ \in [m',m''']$. We see  significant fluctuations around the linear slope $j^{T}_{\mathcal R_1\to {\rm ch}}(m_+) t$,  while for $m_+ \notin [m',m''']$  the fluctuations  are ``negligible''.

  \end{itemize}

The most striking feature in the simulations is undoubtedly the fact that
the current is positive when $m_+<m^{iv}$ so that it flows along the gradient going from the reservoir with smaller magnetization to the one with larger magnetization.  If we dropped the interaction among particles in the channel, namely put $\eps_{x,\ga}\equiv 0$, then the current would flow according to the Fourier law opposite to the gradient, namely from $\mathcal R_2$ to $\mathcal R_1$.

\medskip

{\bf A heuristic argument.}
Let us now imagine to have two channels connected to  $\mathcal R_1$ and $\mathcal R_2$, channel 1 is the channel considered so far while channel 2 is some other channel where the Fourier law is satisfied (for instance the OS-CA with no bias, $\eps_{x,\ga}\equiv 0$, or some
simpler connection as the one discussed later). When  $m_+<m^{iv}$, in channel 1 there is a current $j(m_+)$ going from  $\mathcal R_1$ to $\mathcal R_2$, while in channel 2 the current is $j_2 = \kappa m_+$, $\kappa >0$, going from $\mathcal R_2$ to $\mathcal R_1$ (recall $m_-=-m_+$).  Thus in a time $t$ the reservoir
$\mathcal R_1$ will loose a magnetization $j(m_+)t$ through channel 1 and gain a magnetization $\kappa m_+t$ through channel 2; the opposite happens to $\mathcal R_2$.  This will go forever because the reservoirs in the OS-CA are not changed by what comes and goes; if instead the reservoirs were realized by large but finite systems (as in CC-CA) then after a time which depends on the size of the reservoirs and the difference $j(m_+)- \kappa m_+$ the magnetization in the reservoirs would change and stationarity would be lost.   However if we choose channel 2 so that $\kappa m_+ = j(m_+)$ there is a perfect balance so that what $\mathcal R_1$ gives to $\mathcal R_2$ through   channel 1 comes back from channel 2.  We may thus hope that even if the reservoirs are finite (yet sufficiently large) this is again approximately true and that there is a non zero  current which looks stationary for long times.

The simplest choice for  channel 2 leads to the  CC-CA of Section \ref{sec.j1} where channel 2 is made by just allowing direct exchanges between the two reservoirs.  Then, as we shall see later,
the average  current in the CC-CA from $\mathcal R_2$ to  $\mathcal R_1$ is equal to
$\ga p m_+$, hence the conjecture  that for such a particular value of $\ga p$ there is a non zero stationary current in the circuit which is  close to $j(m_+)$.

\begin{figure}[h!]
\centering
\includegraphics[width=0.70 \textwidth]{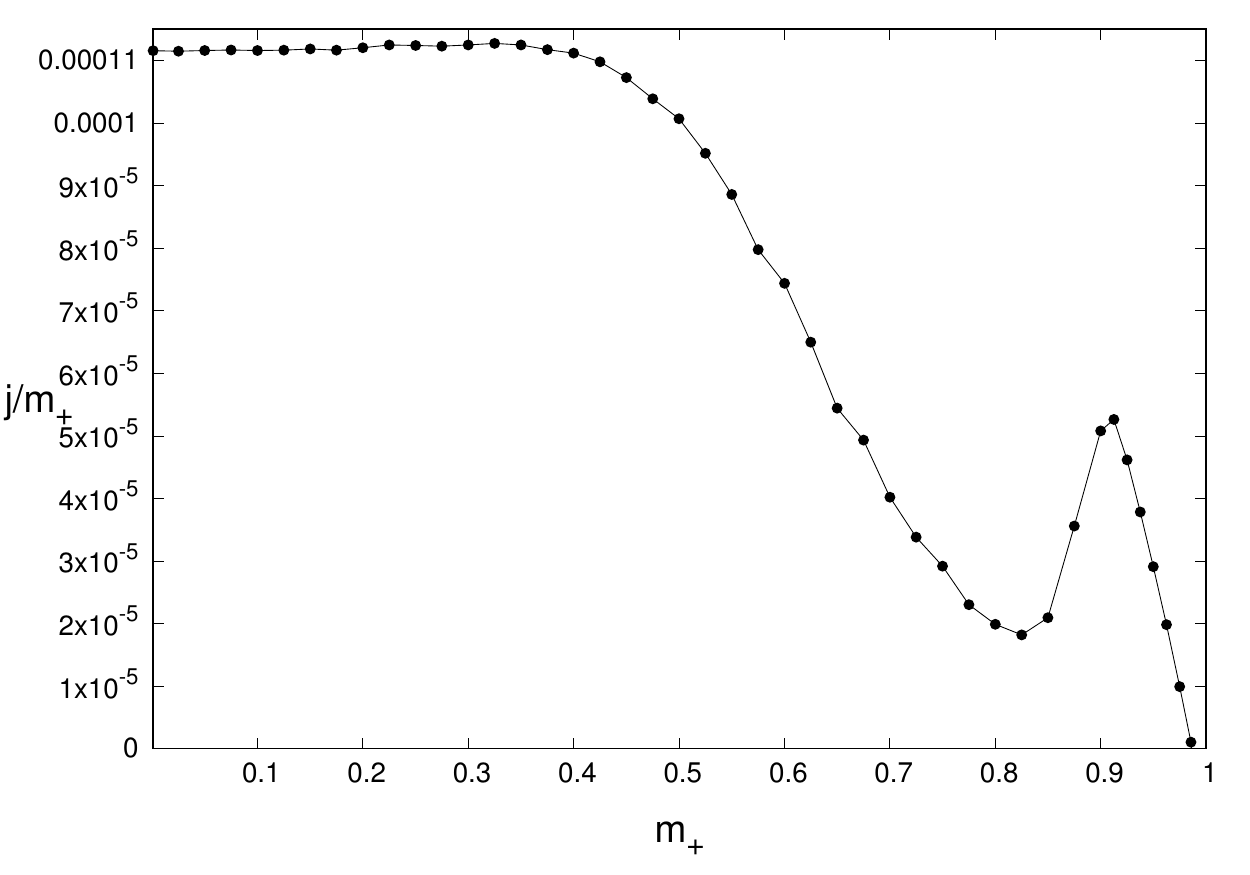}
\caption{{\footnotesize
We plot $j/m_+:=j(m_+)/m_+$ as a function of $m_+$ (black dots). The continuous line is a black dots interpolation.}}
\label{fig:4}
\end{figure}

To check this we have defined for each $m_+<m^{iv}$ in Fig. \ref{fig:2}  $\ga p= j(m_+)/m_+$ as a function of $m_+$, see Fig. \ref{fig:4}.

We have then
run the CC-CA with such values of $\ga p$, putting $m_{+}(0) = m_+$ in $\mathcal R_2$,
$m_{-}(0) = -m_+$ in $\mathcal R_1$ and choosing the initial state in the channel equal to the configuration in the OS-CA simulation at the final time $T$. For all the values of $m_+$ considered in
Fig. \ref{fig:2}
we have run the CC-CA for a same time $T= 3\cdot 10^9$
and computed the averaged currents
$j^{T,CC}_{\mathcal R_1\to {\rm ch}}$,  $j^{T,CC}_{{\rm ch} \to\mathcal R_2}$ and $j^{T,CC}_{\mathcal R_2\to \mathcal R_1}$ defined as  in \eqref{j1.20.2.1}.  
Recalling \eqref{2.5aa} we have also defined the averaged magnetization $m^{T,CC}_{\pm}$ in the two reservoirs writing
$
j^{T,CC}_{\mathcal R_2\to \mathcal R_1}(m_+), j^{T,CC}_{\mathcal R_1\to {\rm ch}}(m_+)$,  $j^{T,CC}_{{\rm ch} \to\mathcal R_2}(m_+), m^{T,CC}_{\pm}(m_+)
$ when we want
to underline that the values are obtained starting from $m_+$.

The previous heuristic argument suggests that the three currents above are all close to each other and thus approximately equal to $j^{T}_{{\rm ch} \to\mathcal R_2}(m_+)$ and moreover
that $m^{T,CC}_{\pm}(m_+)\approx
\pm m_+$.   In the next section we will see  what the simulations say.

\vskip1cm

\setcounter{equation}{0}

\section{Self sustained currents}
\label{sec.j1.1}

\begin{figure}[h]
\centering
\includegraphics[width=0.49 \textwidth]{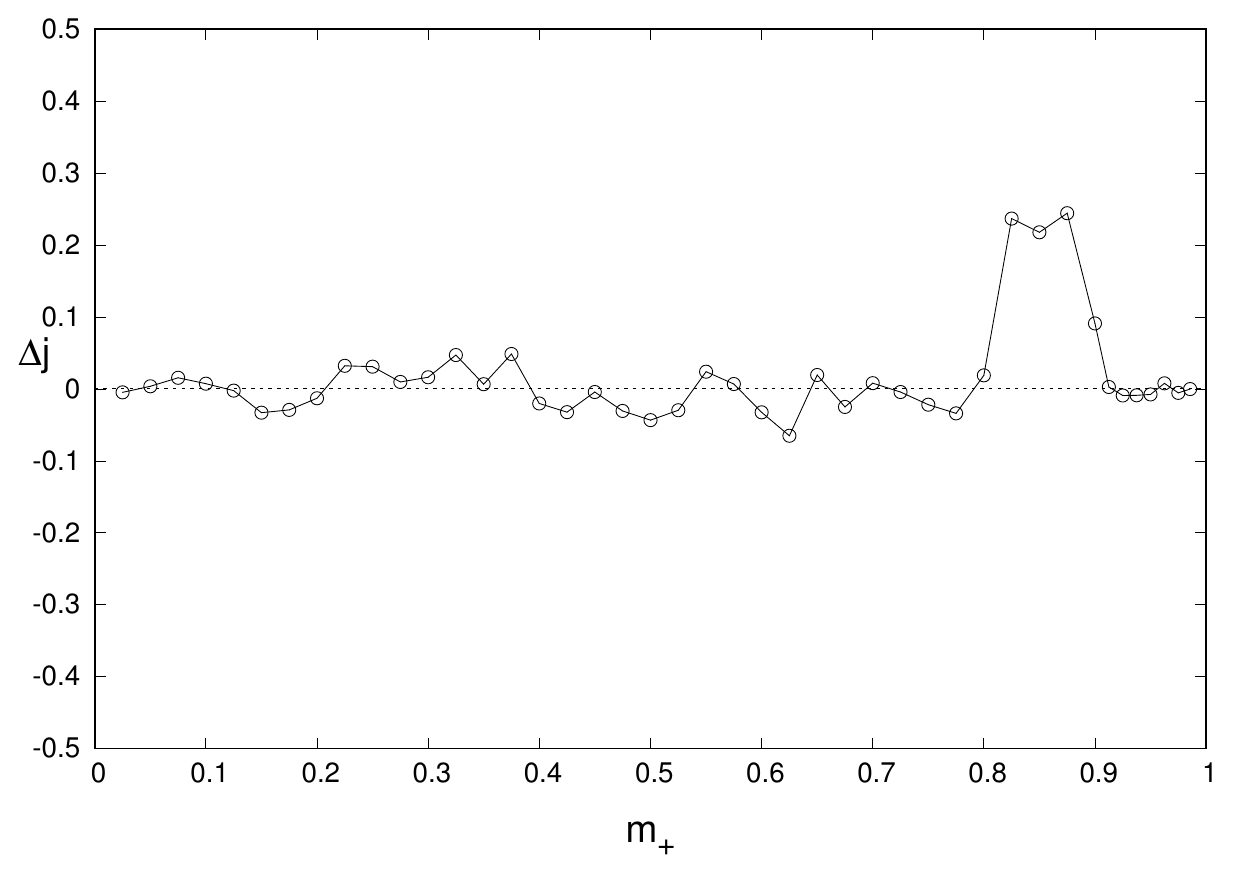}
\hskip 3 pt
\includegraphics[width=0.49 \textwidth]{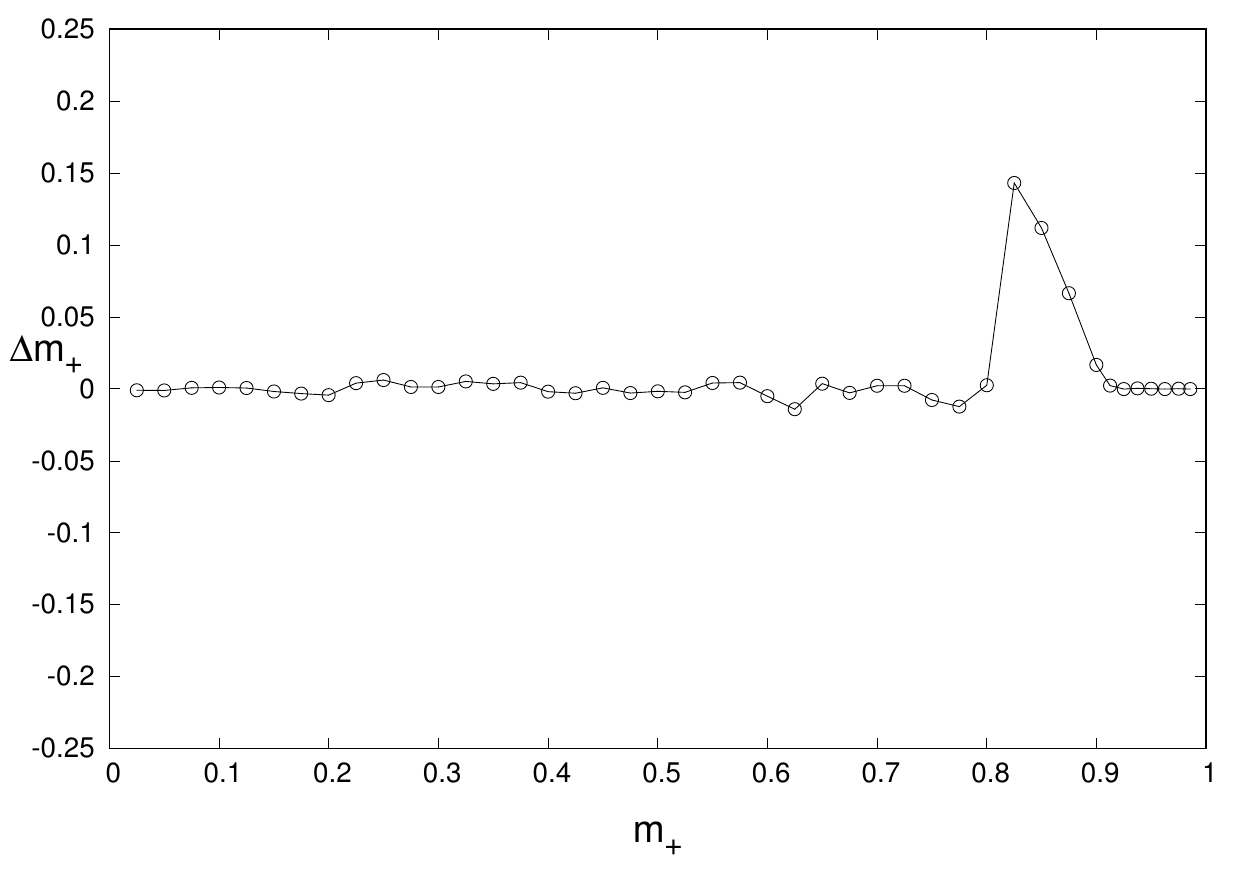}
\caption{{\footnotesize We plot $\Delta j:=10^5\times[j^{T,CC}_{\mathcal R_2\to \mathcal R_1}(m_+)-j^{T,CC}_{{\rm ch} \to\mathcal R_2}(m_+)]$ (left panel) and $\Delta m_+:=m_+^{T,CC}-m_+$ (right panel) as a function of $m_+$. Note that the large fluctuations occur in the interval $(m'',m''')$, with $m''=0.825$, $m'''=0.912$.}}
\label{fig:5}
\end{figure}

Fig. \ref{fig:5} is obtained by running the CC-CA in the setup described at the end of the previous section. It  reports  the values of the differences $10^5[j^{T,CC}_{\mathcal R_2\to \mathcal R_1}(m_+)-j^{T,CC}_{{\rm ch} \to\mathcal R_2}(m_+)]$ and $m_+^{T,CC}(m_+)-m_+$ as a function of $m_+$, recall from Fig. \ref{fig:2} that the typical values of the current have order $10^{-5}$.

\begin{figure}[h]
\centering
\includegraphics[width=0.49 \textwidth]{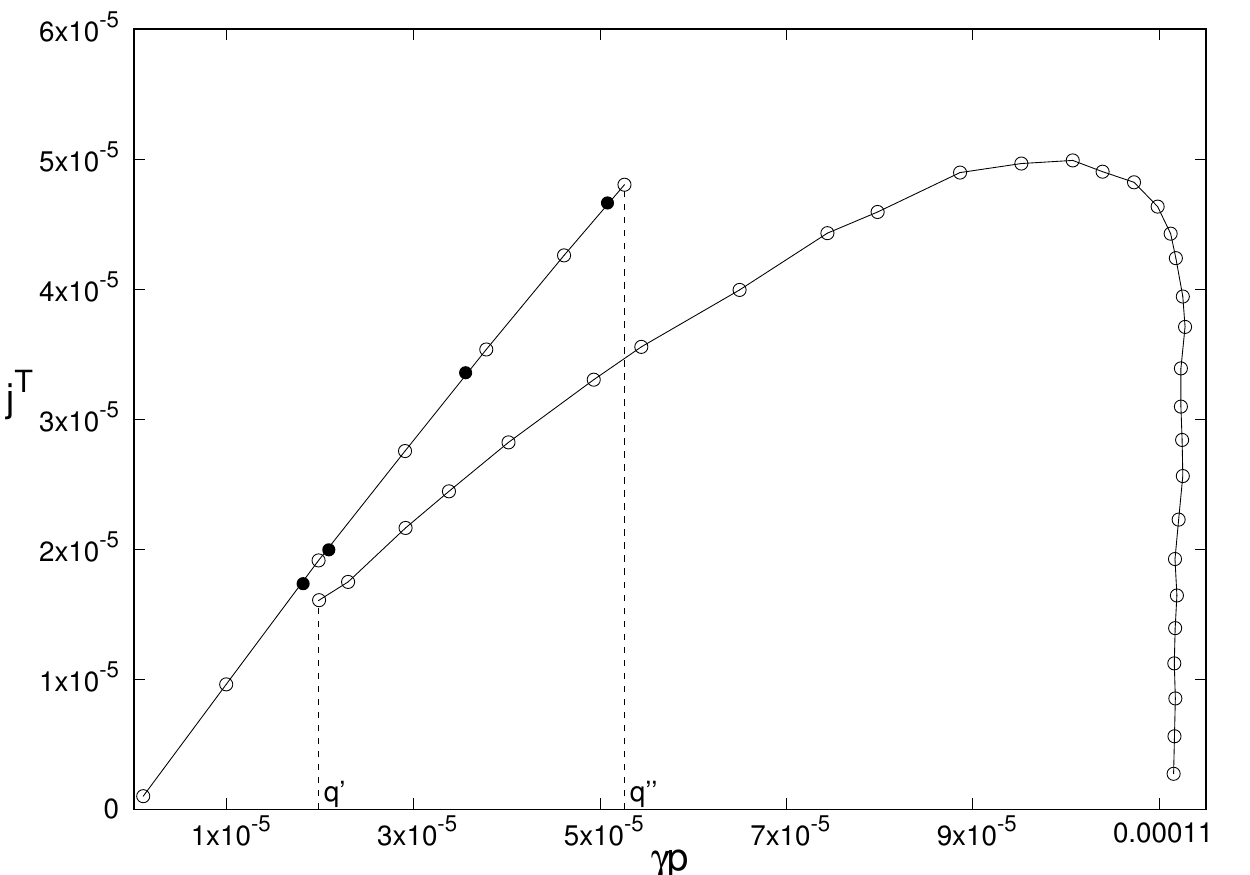}
\hskip 3 pt
\includegraphics[width=0.49 \textwidth]{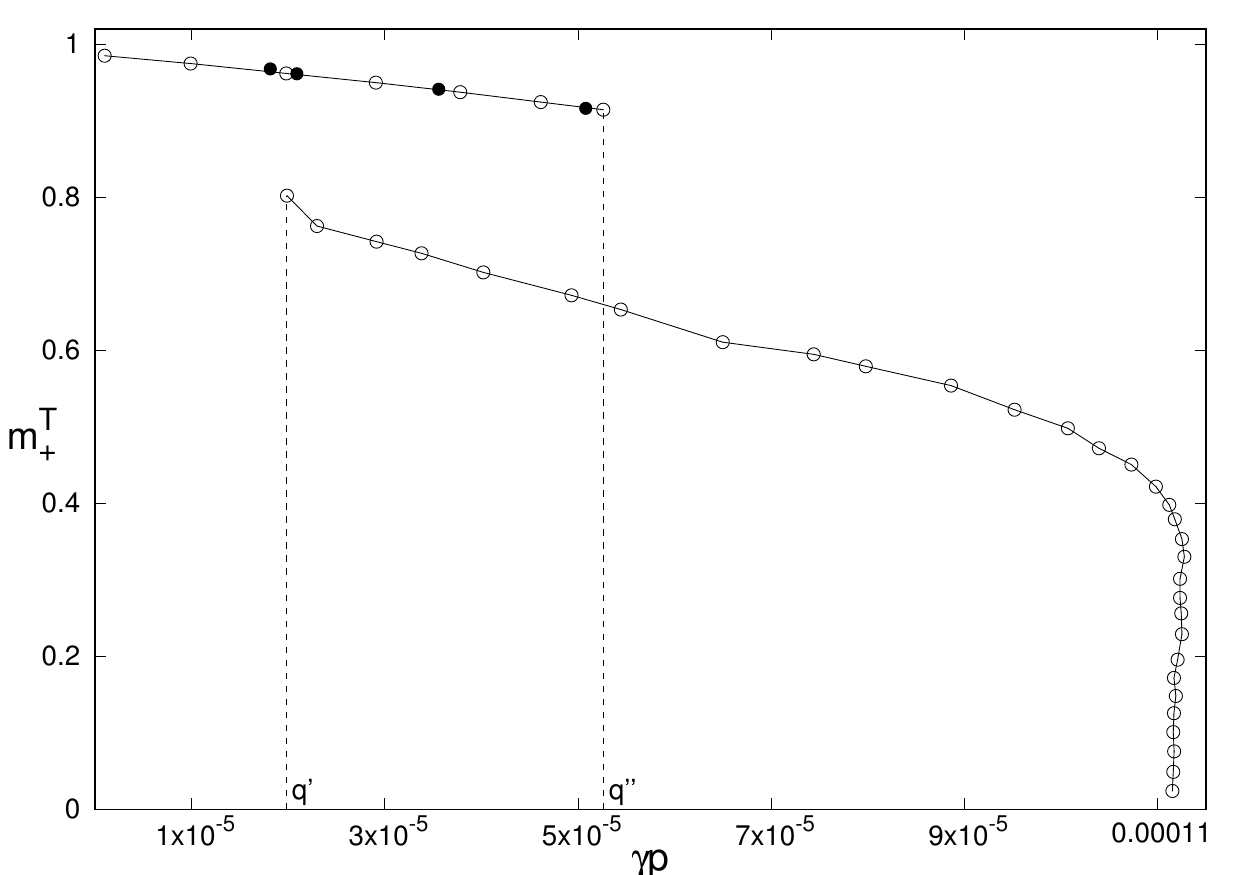}
\caption{{\footnotesize We plot the values of the pairs $(\ga p,j^T:=j^{T,CC}_{\mathcal R_2\to \mathcal R_1})$ (left panel) and $(\ga p,m_+^T:=m_+^{T,CC})$ (right panel). The black circles in the panels above denote,  respectively, the stationary values of  $j^{T,CC}_{\mathcal R_2\to \mathcal R_1}$ and $m_+^{T,CC}$ obtained with $m_+\in (m'',m''')$. Shown are also the values of $q'=1.98\times10^{-5}$ and $q''=5.26\times 10^{-5}$.}}
\label{fig:6}
\end{figure}

We have also reported  for each
$m_+$ in  Fig. \ref{fig:2} the values of the pair
$(\ga p,j^{T,CC}_{\mathcal R_2\to \mathcal R_1})$, see  Fig. \ref{fig:6} left, the continuous line is obtained by interpolating between such values. Analogously in  Fig. \ref{fig:6} right the dots are the values of $(\ga p,m_+^{T,CC})$ and the continuous line is obtained by interpolation.  The continuous lines are multi-valued functions denoted respectively by
$j^{CC}(\ga p)$ and $m_+^{CC}(\ga p)$,  we presume   they are a good approximation of what would be obtained by following the same procedure for other values of $m_+$ in Fig.  \ref{fig:2}.

Let us point out the
 main features of our simulations.

\begin{itemize}

  \item  Fig \ref{fig:5} shows that the simulations
  are in good agreement with the conjectures
  stated at the end of the previous section except in the interval $m_+ \in (m'',m''')$.
  The values of $ j^{T,CC}_{\mathcal R_2\to \mathcal R_1}(m_+)$  and $m_+^{T,CC}(m_+)$ when $m_+ \in (m'',m''')$, are however approximately the same as those obtained for different values of $m_+$, see the black circles in Fig. \ref{fig:6}).

\item
 The values of $\ga p$ are all in the interval $(0,q_c)$, $q_c= 11.25\times10^{-5}$, and $j^{CC}(\ga p)$ is positive for all such values of $\ga p$.  We have also done simulations with $\ga p > q_c$ with several choices of the initial condition and we have always seen zero current (not reported here).
%

    \item
$j^{CC}(\ga p)$ is multi-valued, it has two distinct branches (separated from each other), the upper one
in the interval $(0,q'')$, $q''=5.26\times10^{-5}$, the lower one in the interval $(q',q_c)$; $q''>q'$, $q'=1.98\times10^{-5}$. In the interval $(q',q'')$ there are
two positive currents different from each other. 
   \item
$m_+^{CC}(\ga p)$ has the analogous structure, being two valued in  $(q',q'')$.
 Both branches are decreasing, $m_+^{CC}(\ga p)\to m^{iv} = 0.985$, as
 $\ga p \to 0$, and to $0$ as $\ga p \to q_c$.

    \item There is a gap
 in the range of $m_+^{CC}(\ga p)$, namely
 the interval $(m',m'')$.

\end{itemize}

\medskip

\noindent
{\bf Conclusions.}  The simulations in Fig. \ref{fig:5} show  good agreement
with the conjectures of Section \ref{sec.j1.4} except when $m_+\in (m'',m''')$.  Thus, with such exception, we may say that the stationary state found in the OS-CA evolution persists in the CC-CA provided that  $\ga p=j(m_+)/m_{+}$.

There is no mystery about the current between the two reservoirs being $\ga p(m_{+}-m_-)/2\approx \ga p m_+$ because we can prove (see Appendix \ref{aappA}) that
\begin{equation}
\label{C.0}
 E\Big[ \{j^{T,CC}_{\mathcal R_2\to \mathcal R_1}  - \ga p \frac 12
 [ m_+^{T,CC}-m_-^{T,CC}]\}^2 \Big] \le \frac{\ga p}T +16   \frac {(\ga p)^2}R
 + \; \text{\rm corrections}
\end{equation}
Since $\ga p \approx 10^{-5}$, $R = 10^{5}$ and $T \approx 10^{9}$, the corrections have order $10^{-19}$, see \eqref{C.15}.

Fig. \ref{fig:6} can be obtained from Fig. \ref{fig:2}: in fact according to the above statements $j^{CC}(\ga p)$ is (approximately) equal to $j(m_+)$ with
$j(m_+)= \ga p m_+$.  Since this may have multiple roots, $j^{CC}(\ga p)$ will be correspondingly multi-valued.  However the roots with $m_+\in (m'',m''')$ are absent in the simulations (see the black circles in Fig. \ref{fig:6}) but their values are the same as those obtained with other values of $m_+$.  Same if we look at $m^{CC}(\ga p)$ and compare with Fig. \ref{fig:2}.

As a conclusion we have a consistent explanation of what seen in the OS-CA and the CC-CA, but we still need to explain (i) what happens when $m_+\in (m'',m''')$; (ii) why the typical values of $j(m_+)$ have order $10^{-5}$ which is much smaller than $1/L \approx 10^{-3}$ which is what expected from Fourier law experiments; (iii) why the true reservoir current has the behavior shown in Fig. \ref{fig:2}.

We can gain a theoretical insight on what is going on by looking at what happens in the mesoscopic limit $\ga \to 0$ which we study in the next section.


\setcounter{equation}{0}

\section{The mesoscopic limit}
\label{sec.j1.3}
This is defined by letting $\ga \to 0$ with
\begin{equation}
\label{j1.12}
  L= \ga^{-1} \ell,\quad R = \ga^{-1}a,\quad \ell,a >0\;{\rm fixed}
\end{equation}
In the channel space and time are scaled diffusively,  thus $x \to r =\ga x$ and   $t \to \tau = \ga^{2}t$.
In mesoscopic units the channel after the limit $\ga\to 0$ becomes the real interval $[0,\ell]$.  We will prove existence of the limit (for the relevant quantities)
under the assumption of a strong form of propagation of chaos, the details are given in an appendix.

We denote by $E_\ga$ the expectation in the CA processes (randomness coming from the initial datum
and from the updating rules of the CA's).
\medskip

\noindent {\bf Assumptions}.
 We suppose that
 \begin{enumerate}

\item In both CA the limit below (denoted in the same way for both CA) exists and is smooth
\begin{equation}
	\label{5.3a}
 \lim_{\ga \to 0} \lim_{\ga x\to r, \ga^{2}t\to \tau}  E_\ga[ \eta(x,v,t)] =\frac{m(r,\tau)+1}2, \quad r\in[0,\ell], v\in\{-1,1\}, \tau\ge 0
	\end{equation}

\item In the CC-CA
	\begin{equation}
	\label{j1.15.2}
 \lim_{\ga \to 0, \ga^{2}t\to \tau} m_{\pm,\ga}(\ga^2 t)=
  m_{\pm}(\tau)
	\end{equation}
where, recalling \eqref{2.5aa}, we have set $m_{\pm ,\ga}(\ga^2 t):= E_\ga[m^{CC}_{\pm} (t)]$
%
\item  In both CA  for all $r,r_1,r_2 \in (0,\ell)$, $r_1\ne r_2$, $v\in\{-1,1\}$ and $\tau\ge 0$
	\begin{eqnarray}
\nn
&& \lim_{\ga \to 0} \lim_{\ga x\to r, \ga^{2}t\to \tau} | E_\ga[ \eta(x,v,t)\eta(x,-v,t)]
 - E_\ga[ \eta(x,v,t)]E_\ga[\eta(x,-v,t)] | =0 \\\label{j1.15.3}
 \\&&
\lim_{\ga \to 0} \lim_{\ga x\to r_1, \ga y\to r_2, \ga^{2}t\to \tau} | E_\ga[ \eta(x,t)\eta(y,t)]
 - E_\ga[ \eta(x,t)]E_\ga[\eta(y,t)] | =0
 \nn
\end{eqnarray}
\item   In the CC-CA  for all $\tau\ge 0$
\begin{equation}
\label{j1.15.44}
 \lim_{\ga \to 0, \ga^{2}t\to \tau} R^{-1} E_\ga\Big[ \big|N_{\mathcal R_i}(t)
 -E_\ga[N_{\mathcal R_i}(t)]\big|\Big]=0,\;\; i=1,2
\end{equation}
 \end{enumerate}
\medskip
In Appendix \ref{A1} we will prove the following two Theorems.

\begin{thm} [Mesoscopic limit]
  \label{thm5.1}
Under the above assumptions, in both CA, the limit magnetization $m(r,t)$  satisfies:
	\begin{eqnarray}
		\label{j1.16}
 &&\frac{\partial}{\partial t}m(r,t) =  -
 \frac{\partial}{\partial r} I(r,t),\quad r \in (0,\ell)
 \\&& \hskip-.3cm
 I(r,t)= -\frac 12\big\{\frac{\partial m(r,t)}{\partial r}
 - 2C[1-
 m(r,t)^2]\int_{r}^{r+1}[m(r+\xi,t)
 - m(r-\xi,t)] d\xi\big\}\nn
 \end{eqnarray}
with $m(r+\xi,t)= m_+(t)$ if $r+\xi\ge \ell$ and $m(r-\xi,t)= m_-(t)$ if $r-\xi\le 0$ in the CC-CA; same expression holds in the OS-CA but with $m_\pm(t)$ replaced by $m_\pm$.  Moreover
	\begin{equation}
	\label{5.7}
 m(0,t)=m_{-},\quad m(\ell,t)=m_{+},\qquad \text{in the OS-CA}
	\end{equation}
while in the CC-CA
	\begin{equation}
	\label{j1.17}
 m(0,t)=m_{-}(t),\quad m(\ell,t)=m_{+}(t) 
	\end{equation}
\begin{equation}
\label{j1.18}
\frac{d}{dt}m_{+}(t) = \frac 1a \Big(2I(\ell, t)+  p [m_{-}(t)-m_{+}(t)]\Big)
\end{equation}
\begin{equation}
\label{j1.19}
\frac{d}{d t}m_{-}(t) = \frac 1a \Big( -2I(0,t)+   p [m_{+}(t)-m_{-}(t)]\Big)
\end{equation}

\end{thm}

\bigskip

 A proof  which avoids our assumptions of propagation of chaos has been obtained in \cite{GL} for a lattice gas with Kac potential  and Kawasaki dynamics in a torus. In magnetization variables the system becomes the Ising model with Kac potential and the limit equation is \eqref{j1.16}. In  \cite{LOP}  it has been studied the macroscopic scaling limit of this system with space scaled by $\ga^{-\alpha}$ and time by $\ga^{-2\alpha}$, $\alpha>1$ ( $\alpha =1$ is the mesoscopic limit considered above).

\begin{thm} [Currents]
  \label{thm5.1b}
Denote by $j_{x,x+1}(t)$ the number of particles which in the time step $t, t+1$  cross the bond $(x,x+1)$, $x\in\{1,..,L-1\}$ (counting as positive those which jump from $x$ to $x+1$ and as negative those from $x+1$ to $x$).  Then, under the above assumptions, in both CA,  for all $r\in(0,\ell)$ and $\tau>0$
\begin{equation}
\label{B.12}
 \lim_{\ga\to 0: \ga x\to r} \ga \sum_{t=0}^{T-1} E_\ga[j_{x,x+1}(t)]= \int_0^\tau I(r,s)ds,\qquad T=[\ga^{-2}\tau]
 	 \end{equation}
						\begin{eqnarray}
\label{j1.19.1}
&&\lim_{\ga\to 0}\ga \sum_{t=0}^{T-1} E_\ga[ j_{\mathcal R_1\to {\rm ch}}(t)]= -
\int_{0}^{\tau} I(0,s)ds,\nn\\&&
\lim_{\ga\to 0}\ga \sum_{t=0}^{T-1} E_\ga[j_{{\rm ch}\to \mathcal R_2}(t)]=
\int_{0}^{\tau} I(\ell,s)ds
		\end{eqnarray}
where $I(r,s)$ is given in \eqref{j1.16}.
\end{thm}

\medskip

In the CC-CA the current between reservoirs converges
by \eqref{C.0} to:
\begin{eqnarray}
\label{j1.19.1.1}
&&\lim_{\ga\to 0}\ga \sum_{t=0}^{T-1} E_\ga[ j_{\mathcal R_2\to \mathcal R_1}(t)]=
\int_{0}^{\tau}  p m_+(s)ds
\end{eqnarray}

In the simulations we have plotted the quantity $ j^{T}_{{\rm ch}\to \mathcal R_2}$.  This is related by \eqref{j1.19.1}   to the mesoscopic current $I$ by
	\begin{eqnarray}
	\label{5.12a}
&&E_\ga[ j^{T}_{\mathcal R_1\to {\rm ch}}]=\frac{\ga}{T} \sum_{t=0}^{T-1} E_\ga[ j_{\mathcal R_1\to {\rm ch}}(t)] \approx - \frac {\ga}{\tau}
\int_{0}^{\tau} I(0,s)ds	,\nn\\&&
E_\ga[ j^{T}_{{\rm ch}\to \mathcal R_2}]=\frac{\ga}{T} \sum_{t=0}^{T-1} E_\ga[j_{{\rm ch}\to \mathcal R_2}(t)] \approx \frac {\ga}{\tau}
\int_{0}^{\tau} I(\ell,s)ds	
\nn\\&&
E_\ga[ j^T_{\mathcal R_2\to \mathcal R_1}]=
\frac{\ga}{T} \sum_{t=0}^{T-1} E_\ga[ j_{\mathcal R_2\to \mathcal R_1}(t)]
\approx \frac {\ga}{\tau} 
\int_{0}^{\tau}  p m_+(s)ds
		\end{eqnarray}
so that the experimental values of the three currents scale all as $\ga$ when $\ga\to 0$.

 We next show that there is a natural interpretation of the solutions of the system \eqref{j1.16}--\eqref{j1.19} in terms of statistical mechanics,
    which then allows to relate  what seen in the simulations to phase transitions and metastable--unstable magnetization values.

%

\medskip

{\bf Free energy functional and thermodynamic potentials.}
The evolution equation \eqref{j1.16} in $[0,\ell]$ with periodic boundary conditions is the gradient flow relative to a non local free energy functional $F(m)$, in fact
\begin{eqnarray}
\label{j1.23}
 &&
 I(r)= - \chi \frac {\partial}{\partial r}\frac {\delta F(m)}{\delta m(r)},\quad \chi =\frac\beta 2(1-m^2), \quad  \beta=2C
 \\&& F(m) = \int \Big(-\frac {m^2}{2} - \frac {S}{\beta}\Big) + \frac 14\int\int
 J (r,r') [m(r)-m(r')]^2
 \nn
 \\&&S(m)= -\frac{1-m} {2}\log \frac{1-m} {2} - \frac{1+m} {2}\log \frac{1+m} {2}
 \nn
 \\&& J(r,r')=1-|r-r'|,\qquad \text {for }  |r-r'|\le 1\quad \text{ and }=0 \quad \text {elsewhere}
\nn
\end{eqnarray}
$F(m)$ is
``the mesoscopic free energy functional'', the Ginzburg-Landau functional is a local
approximation of $F(m)$ where the non local term becomes a gradient squared.  The corresponding gradient flow evolution is the Cahn-Hilliard equation, which can then
be viewed as a local approximation of \eqref{j1.16}.

The important point for us is that $F(m)$ specifies the thermodynamics of the system.  In fact
\begin{eqnarray}
\label{j1.24}
 &&
 f_\beta(m)= -\frac {m^2}{2} - \frac {S(m)}{\beta}
\end{eqnarray}
 is the van der Waals mean field free energy; its convex envelope $f^{**}_\beta(m)$ is the thermodynamic free energy. $f^{**}_\beta(s)$ is obtained  by minimizing $F(m)/\ell$ under the constraint $\int m(r) = \ell s$ and then taking the limit $\ell \to \infty$, see for instance \cite{presutti}, Ch. 6.

 The equilibrium magnetization density when there is a magnetic field $h$ is the solution of the mean field equation
\begin{eqnarray}
\label{j1.25}
 &&
  m =  \tanh\{\beta (m+h)\}
\end{eqnarray}
When $\beta>1$ there is $h_c(\beta)>0$ so that for any $|h|<h_c(\beta)$,
$f_\beta(m,h)=f_\beta(m)-hm $ is a double well function of $m$.  The local minima are
$m_+(h)$ and $m_-(h)$ and their graph is the hysteresis cycle, see Fig.  \ref{fig:1}.
In particular at $h=-h_c(\beta)$, $m_+(h)= m^*$
\begin{eqnarray}
\label{j1.26}
 &&
  m^*>0 :  \beta [1-(m^*)^2] =1
\end{eqnarray}
so that the magnetization in $(m^*,m_\beta)$ and in $(-m_\beta,- m^*)$ is metastable.
At $h=0$ the double well is symmetric and the local minima are global minima,
they are attained at $m=\pm m_\beta$, $m_\beta $ the positive solution of \eqref{j1.25} with $h=0$. $\pm m_\beta$ are the equilibrium magnetization at the phase transition with $h=0$ and $\beta>1$.  $m_+(h)$ and  $m_-(h)$
are the unique equilibrium magnetization at   $h> 0$ and respectively $h<0$.

\medskip
%

\vskip1cm
\section{The adiabatic limit.}  
\label{sec.j1.5}
Some of the characteristic  parameters of the
simulations are related to the thermodynamics associated to the mesoscopic equations, see the end of Section \ref{sec.j1.3}.
%
%
%
Indeed in fig \ref{fig:2} which refers to simulations with the OS-CA, the value
$ 0.985$ is very close to $m_\beta$ so that the simulation shows that the current is negative when $m_+$ is stable, namely $m_+ > m_\beta$ and positive when $m_+ < m_\beta$ (metastable or unstable). Correspondingly when there is a current in the CC-CA then $m_+ < m_\beta$, see fig. \ref{fig:6} right.
The above validates the considerations in the Introduction about the relation between the appearance of a current in the circuit and the occurrence of phase transitions.

Also the metastable region $(m^*,m_\beta)$ has a  role in the simulations as
the interval $(m'',m''')$ is a subset of $(m^*,m_\beta)$ (because  $m''=0.825$
and $m'''=0.912$ while $m^*\approx 0.775$ and $m_\beta \approx 0.985$); thus the
gap phenomenon (i.e.\  that some values of the magnetization in $\mathcal R_2$ are never
seen
for all $\ga p$) occurs only inside the metastable region.
%
%

We turn now to the heuristic argument at the end of Section \ref{sec.j1.4} by observing that it becomes rigorous in the context of the mesoscopic equations. In fact if $m$ is a stationary solution of the mesoscopic equation for the OS-CA when the reservoirs magnetizations are $m_\pm$, and the corresponding current $I$ is positive,
then  $u=m$ in the channel and $u=m_\pm$  in $\mathcal R_i$ $i=1,2$ is a stationary solution of the CC-CA mesoscopic equations with $p= \frac I{\frac 12 (m_+-m_-)}$, recall that
the ratio between  the mesoscopic current and the current in the CA scales as $\ga^{-1}$.  Thus for sufficiently small $\ga$ we may expect to see what conjectured at the end of Section \ref{sec.j1.4}.

We can also give an explanation of the gap phenomenon  (i.e.\  that for all $\ga p$ the magnetization in the reservoirs $\mathcal R_2$ is never in the interval  $(m'',m''')$)
by assuming that the evolution of the OS-CA is well approximated by the mesoscopic equations in the adiabatic limit that we are going to define.  We first observe that
the OS-CA can be regarded as the ``infinite reservoirs limit'' of the CC-CA, in fact in the limit $R\to\infty$ the updating rules of the CC-CA become those of the OS-CA. This is true also at the mesoscopic level: when $a\to\infty$ the magnetizations $m_\pm(t)$ converge to their initial value $m_\pm(0)$ and the evolution becomes that of the OS-CA.
The above is true when we let $a\to \infty$ keeping the time finite,
more interesting behaviour is seen if we scale time proportionally to $a$, which is the so called adiabatic scaling limit. Suppose (in agreement with the simulations in fig. \ref{fig:2}) that for each value of $m_+$ (and with $m_-=-m_+$) there is a unique stationary solution of the mesoscopic equations for the OS-CA,
 $I_{\rm{stat}}(m_+)$ being the corresponding current.
We then say that the CC-CA mesoscopic equations   have a ``good adiabatic behavior'' if in the adiabatic limit
the magnetizations $m_\pm(t)$  satisfy the equations
	\begin{eqnarray}
 \label{j4.1}
  \frac{d m_{+}(t)}{d t} = 2\Big( I_{\rm{stat}}(m_{+}(t)) -  pm_{+}(t))
 \Big) \qquad m_-(t)=-m_+(t)
	\end{eqnarray}
Suppose now that $ I_{\rm{stat}}$  is positive with  a graph like $j(m_+)$, see fig. \ref{fig:2}. Then the stationary solutions of $p m_+= I_{\rm{stat}}(m_+)$ with $m_+\in(m'',m''')$ are linearly unstable because $I_{\rm{stat}}(m_+)$ is decreasing while $p m_+$ is increasing.  Thus a small perturbation will lead the magnetization away from the stationary value $p m_+= I_{\rm{stat}}(m_+)$, $m_+\in(m'',m''')$, and presumably it will converge to one of the two other solutions of $p m_+= I_{\rm{stat}}(m_+)$.  This
may therefore explain why in the simulations we do not see the magnetization $m_+\in(m'',m''')$ and instead find another solution of $p m_+= I_{\rm{stat}}(m_+)$.

\begin{figure}[h]
\centering
\includegraphics[width=0.49 \textwidth]{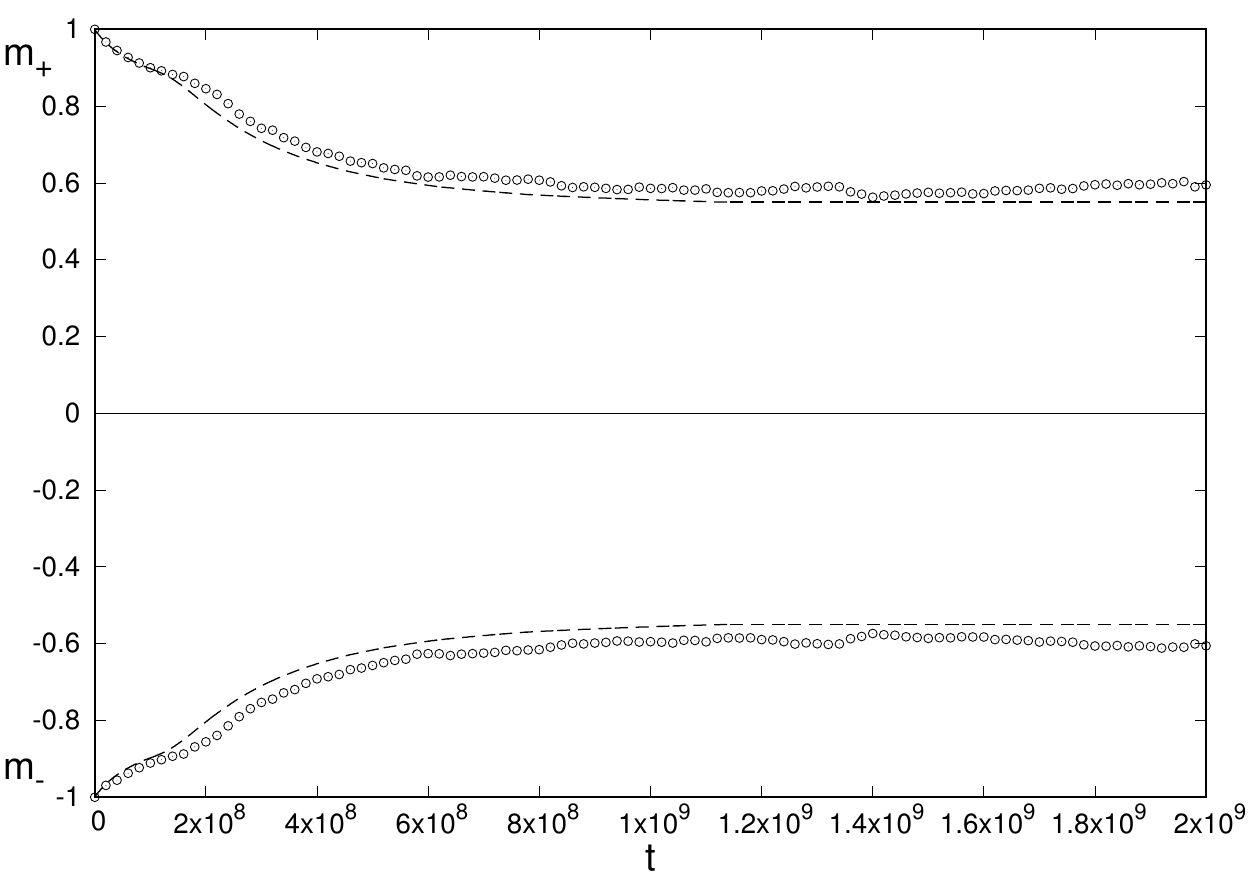}
\hskip 3pt
\includegraphics[width=0.49 \textwidth]{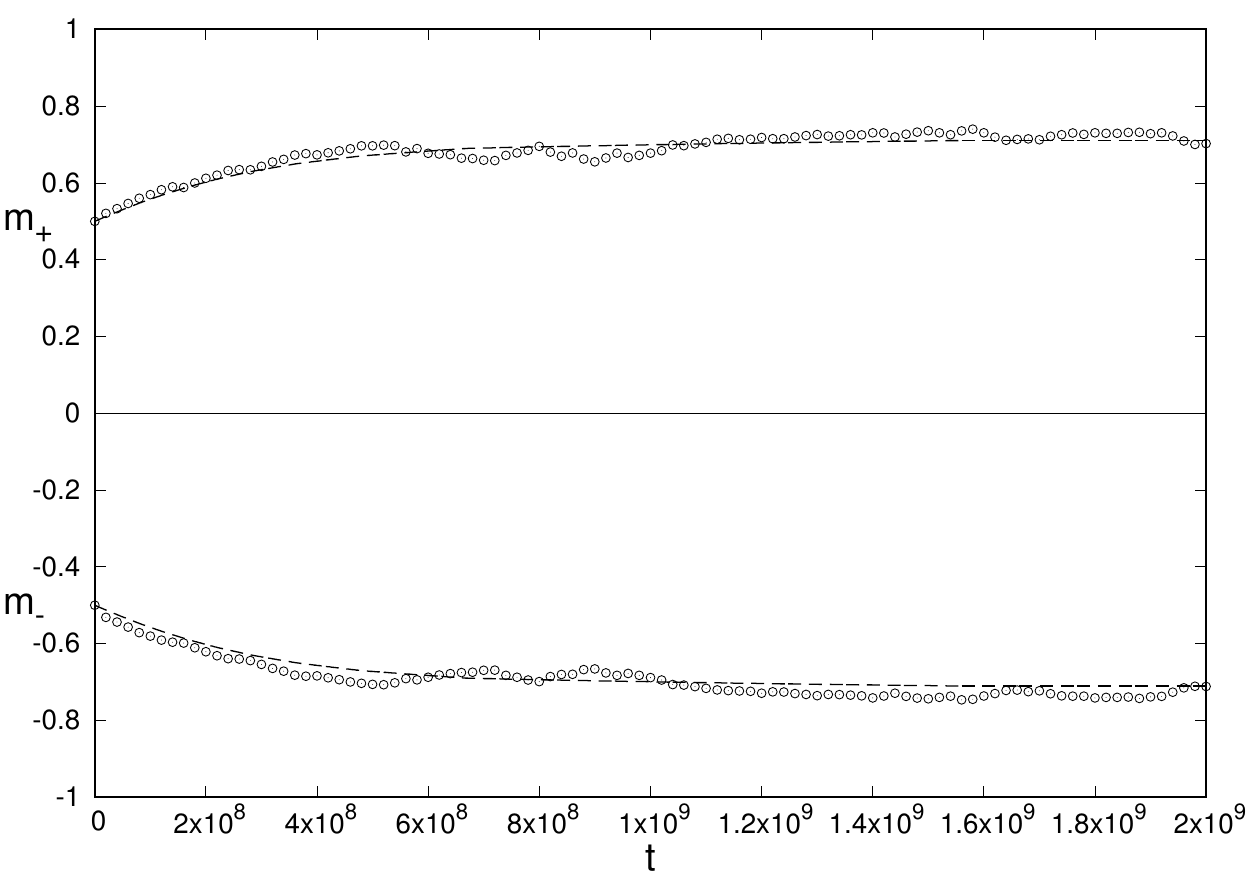}
\caption{{\footnotesize We plot $m_{\pm}$ as functions of time $t$ (empty circles), obtained by running the CC-CA, as well as the predicted behavior in the adiabatic limit (dashed line). The initial values of the magnetization are, respectively, $m_+(0)=-m_-(0)=1$ (left panel) and $m_+(0)=-m_-(0)=0.5$ (right panel).}}
\label{fig:adia}
\end{figure}

We can check experimentally whether the CC-CA has a good adiabatic behavior by doing
simulations with non stationary initial data.  In fig. \ref{fig:adia} we report the experimental values and those obtained by solving numerically the adiabatic equations.

We do not have an analytic proof of good adiabatic behavior which instead can be rigorously proved for another particle model.  This is the simple symmetric exclusion process in
an interval with boundary processes at the endpoints which simulate reservoirs with densities $\rho_{\pm}(t)$ dependent on time.  In \cite{DO} it is proved that in a
scaling limit where
$\rho_{\pm}(t)$ are ``slowly varying'' the current in the system becomes at each time $t$ the same as the stationary current when the densities at the endpoints are kept fixed at the values $\rho_{\pm}(t)$.

Summarizing, we have a reasonable explanation of the simulations in the CC-CA once we accept
the behavior of the current $j(m_+)$ in the OS-CA as given in Fig. \ref{fig:2}.  To explain the latter we need to go deeper in the analysis of the simulations discussing the magnetization profile in the channel, which will be the argument of the remaining sections.

%
%
%
%
%
%
%
%
%
%
%
%
%

\vskip1cm

\setcounter{equation}{0}
\section{The instanton and the Stefan problem}
	\label{sec6}
\begin{figure}[h]
\centering
\includegraphics[width=0.70 \textwidth]{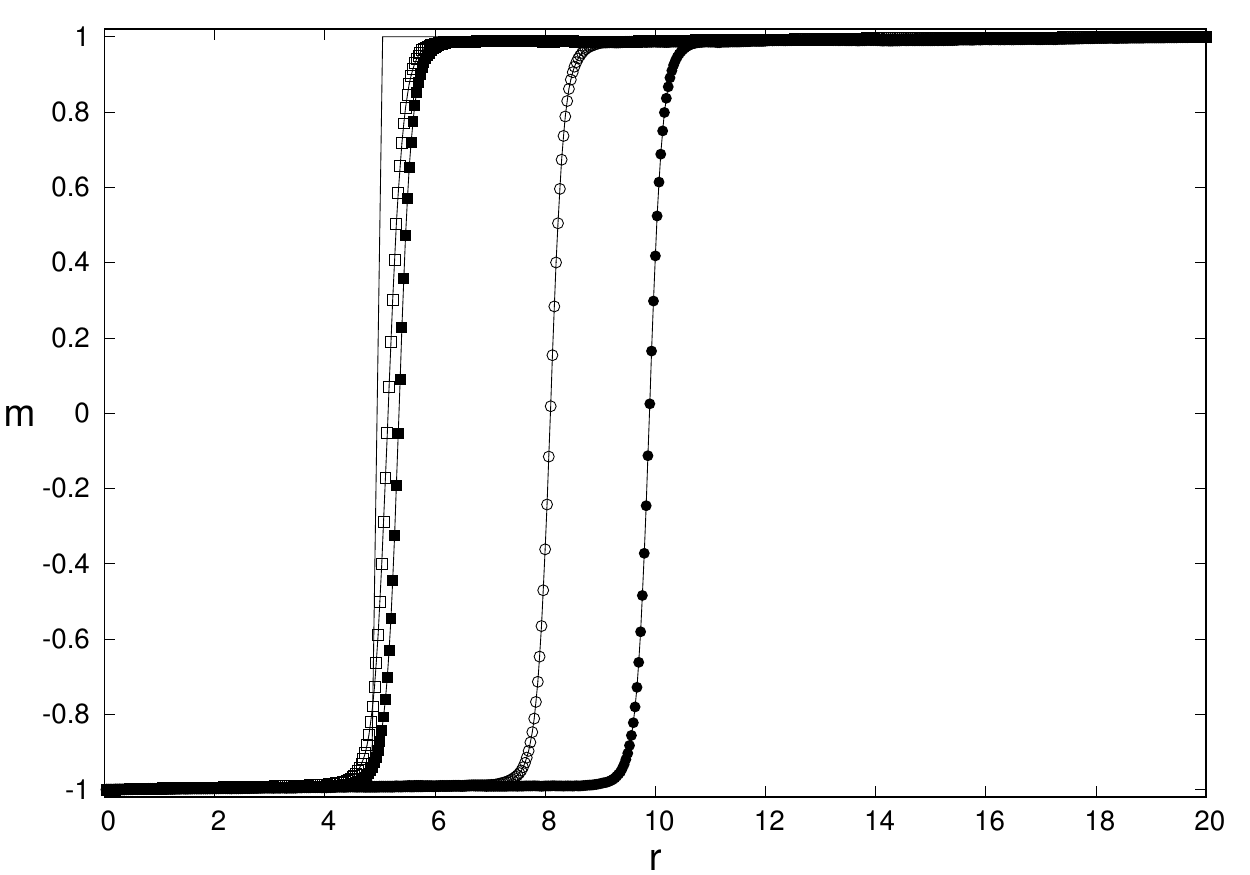}
\caption{{\footnotesize Magnetization profile with $m_{+}=1$. The different curves in the plot correspond to the averaged magnetization computed at different times: $t=10^5$ (empty squares), $t=10^6$ (black squares), $t=10^7$ (empty circles) and $t=10^8$ (black circles). The black thin line denotes the initial configuration, corresponding to a step function centered at $r=5$.}}
\label{fig:j3.2}
\end{figure}

We have a good understanding of what happens when $m_+\in (m_\beta,1]$.  In Fig. \ref{fig:j3.2} we plot the time evolution of the magnetization pattern when $m_+ = 1$, but a
similar picture is observed for the other values of $m\in (m_\beta,1]$. The simulation shows convergence as time increases to a profile which is therefore stationary (in the times of the simulation) and it agrees with what found studying the mesoscopic equations.
The existence of stationary solutions $m_{\rm st}(r;\ell;m_{\pm})$ of \eqref{j1.16} with boundary conditions \eqref{5.7} when $m_+>m_\beta$ has been proved
in \cite{DPT} for $\ell$ large enough.  It is also shown that
 \begin{equation}
\label{j3.5}
\lim_{\ell \to \infty}m_{\rm st}(r\ell;\ell;m_{\pm})=m_{\rm st}(r;m_{\pm}),\quad r\in (0,1)
\end{equation}
where the limit $m_{\rm st}(r;m_{\pm})$ is antisymmetric around $r= 1/2$ and satisfies the equation
 \begin{equation}
\label{j3.6}
-\frac 12 [1-\beta(1-m_{\rm st}^2)]\frac{dm_{\rm st}}{dr} =I_{\rm st}(m_+),\quad r\in[\frac 12,1]
\end{equation}
where $I_{\rm st}(m_+)$ is determined by requiring that $m_{\rm st}(1/2)=m_\beta$ and
$m_{\rm st}(1)=m_+$.
For $m_+=1$, $\beta=2.5$ and $m_\beta=0.985$ from \eqref{j3.6} we get $I_{\rm st}(1)\simeq -7.2\times 10^{-3}$. To compare with the simulations we have to divide
by $L=\ga^{-1}\ell= 600$ getting $-1.2 \times 10^{-5}$  the current in the simulations of the cellular automata OS -CA  is instead $\simeq -2.2\times 10^{-5}$.  The discrepancy is possibly due to $\ell$ not being large enough. In \cite{DPT} it is also proved that
 \begin{equation}
\label{j3.6.1}
\lim_{\ell \to \infty}m_{\rm st}(\frac 12 + x;\ell)=\bar m (x),\quad x\in \mathbb R
\end{equation}
where $\bar m (x)$ is the instanton solution of
 \begin{equation}
\label{j3.6.2}
\bar m (x) = \tanh\{J*\bar m (x)\}
\end{equation}
namely the antisymmetric function solution of \eqref{j3.6.2}
which converges to $m_\beta$ as $x \to \infty$.  See for instance \cite{presutti} for existence and properties of the instanton.

In Fig. \ref{fig:j3.2} it is also plotted the time evolution of the magnetization pattern when starting away from the stationary one.  The approach to the latter occurs on the time scale $L^2$.

\medskip
{\bf Conjecture.} {\em  Let $m(r,t;\ell;m_{\pm})$ be the solution of  \eqref{j1.16} with boundary conditions \eqref{5.7} and with initial datum  $m_0(r\ell)$, $r\in [0,1]$,  such that:

\begin{itemize}

\item
$m_0(r) < -m_\beta$ is smooth in $r<r_0$, $r_0 \in (0,1)$
with limits $m_-$ and $-m_\beta$ as $r\to 0$ and $r\to r_0$

\item $m_0(r) >m_\beta$ is smooth in $r>r_0$,
with limits $m_\beta$ and $m_+$   as $r\to r_0$ and $r\to 1$.

\end{itemize}

Then
 \begin{equation}
\label{j3.7}
\lim_{\ell \to \infty} m(r\ell,t\ell^2;\ell)= m(r,t)
\end{equation}
where $m(r,t)$ is the solution of the Stefan problem with initial datum $m_0(r)$:
\begin{equation}
\label{j3.8}
 \frac{\partial}{\partial t}m(r,t) =  -
 \frac{\partial}{\partial r} I(r,t),\quad
I(r,t)= -\frac 12 [1- \beta (1- m(r,t)^2)]\frac{\partial}{\partial r}m(r,t)
 \end{equation}
where \eqref{j3.8} holds in $\{r<r_t\}$ and in  $\{r>r_t\}$ with Dirichlet boundary conditions $m_-$ and $-m_\beta$ in  $\{r<r_t\}$ and $m_\beta$ and $m_+$ in $\{r>r_t\}$.  The free boundary $r_t$ is also an unknown and it is determined by
\eqref{j3.8} and  the condition }
\begin{equation}
\label{j3.9}
 2m_\beta\frac{dr_t}{dt}  =  I(r_t^-,t) - I(r_t^+,t)
 \end{equation}

 \vskip.5cm

We do not have a proof that $m(r,t)\to m_{\rm st}(r;m_{\pm})$ as $t\to \infty$ ($m_{\rm st}(r;m_{\pm})$ as in \eqref{j3.5}).  However if the pattern  looks like the one in Fig. \ref{fig:j3.2}, i.e.\ essentially linear away from $\pm m_\beta$,  then the current (being proportionally to the slope) when $r> r_t > 1/2$ is larger (in absolute value) than the one when  $r< r_t$.  Thus the magnetization increases and therefore $r_t$ moves to the left.

\eqref{j3.8} has been derived in \cite{LOP} from the spin dynamics on a torus when the initial profile $m_0(r)$ has values in $(m^*,1)$ for all $r$ or when it has values in $(-1,-m^*)$.  The result does not apply in the case of the Stefan problem where there are both positive and negative values of the magnetization: the derivation of the Stefan problem for Ising spins with Kawasaki dynamics and Kac potential is still an open problem.

\section{Boundary layers, the bump}
\label{sec7}

\begin{figure}[h]
\centering
\includegraphics[width=0.70 \textwidth]{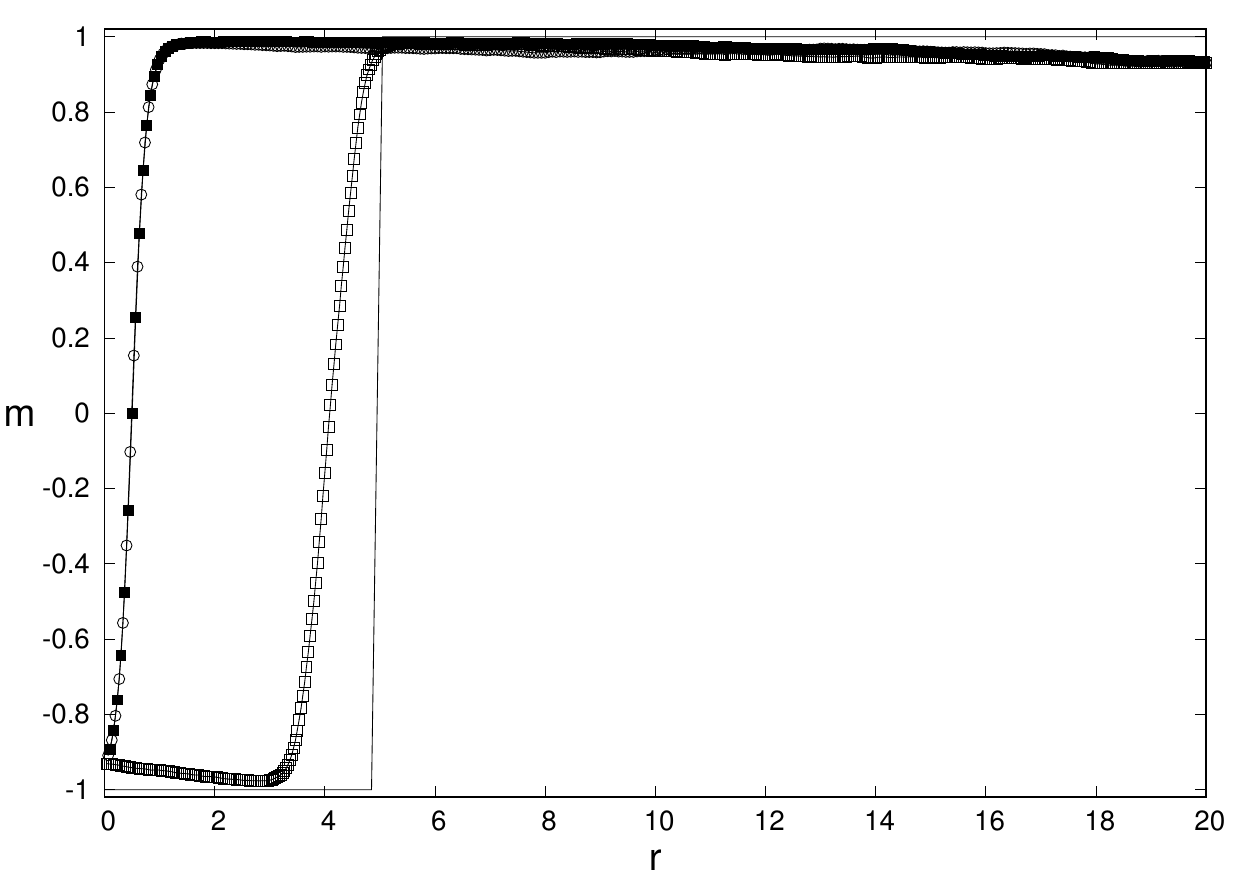}
\caption{{\footnotesize Magnetization profile with $m_{+}=0.93$. The different curves in the plot correspond to the averaged magnetization computed at different times: $t=10^5$ (empty squares), $t=10^6$ (black squares) and $t=10^8$ (empty circles). The black thin line denotes the initial configuration, corresponding to a step function centered at $r=5$.}}
\label{fig:j3.3}
\end{figure}

When $m_+<m_\beta$ we get a completely different picture.  Compare in fact the simulations in
Fig. \ref{fig:j3.2} and Fig. \ref{fig:j3.3} where the initial state is the same but $m_+$ is stable in the former ($m_+=1$) and metastable ($m_+=0.93$) in the latter. In both cases, after a transient, we see 
a  profile with a sharp (instanton-like) transition from $-m_\beta$ to $+m_\beta$ and then approximately
linear profiles which connect $m_-$ to $-m_\beta$ and $m_\beta$ to $m_+$. But, in the stable case the instanton-like region  moves towards the center, while  in the metastable case it moves  towards 0 which is eventually reached.  
The same [heuristic] argument which explained in the case of Fig. \ref{fig:j3.2} the motion of the instanton towards the center, now explains its motion away from the center: since $m_+<m_\beta$   the slope of the pattern from the endpoint to the instanton is  negative in the case of Fig. \ref{fig:j3.3} (as it connects $m_-$ to $-m_\beta$ and $m_\beta$ to $m_+$);
consequently the current in the interval from $m_-$ to $-m_\beta$ is positive and larger
than the one from $m_\beta$ to $m_+$ (as the instanton in Fig. \ref{fig:j3.3} is closer to 0 than to $\ell$), thus the total magnetization increases and the instanton moves further  towards 0.

The transition region in the stable case is approximated by
an instanton which is a stationary solution of the evolution equations on the whole line.  Analogously, when $m_+<m_\beta$ we speculate that the transition region is approximated by a bump which is again a stationary solution $m(r)$, $r\ge 0$,
of \eqref{j1.16} on the half line with zero current and given boundary condition at 0, say $\mu$, namely:
\begin{eqnarray}
\label{j3.11}
 && m(r) = \tanh\Big\{ \beta[ J*m(r)+h]\Big\},\quad r \ge 0
 \\&& h= -J*m(0) + \frac 1 \beta\tanh^{-1} \mu, \qquad m(r)=\mu \,\,\text{for $r<0$} \label{j3.11bis}
 \end{eqnarray}
Indeed it can be easily seen that
a stationary solution of \eqref{j1.16} with zero current is necessarily a solution  of \eqref{j3.11}. The Gibbsian formula (in the mesoscopic limit) would give \eqref{j3.11} with $h=0$, thus the problem \eqref{j3.11} is not in the framework of the equilibrium theory.  This is reflected by the appearance of an auxiliary magnetic field which has to be determined consistently with the magnetization pattern (as in the  FitzHugh Nagumo models of the introduction where however by a mean field assumption the magnetization was simply
a real number).

Observe that if $m(r)$ solves  \eqref{j3.11} with boundary condition $\mu$
then $-m(r)$ solves  \eqref{j3.11} with boundary condition $-\mu$, this symmetry will play an important role in the sequel.  Besides
the trivial solution $m(r)\equiv \mu$,   existence of other solutions of  \eqref{j3.11} is an open problem. The simulations indicate the existence of increasing solutions, we thus define:

\medskip

{\bf Definition.} {\em  The bump $B_{\mu}(r)$, $\mu \in (-m_\beta, m^{*})$,  is a non constant solution of  \eqref{j3.11} which is monotone non decreasing, We call $b(\mu)$ its asymptotic value:}
\begin{equation}
\label{j3.12}
 \lim_{r \to \infty}  B_{\mu}(r)=: b(\mu)
 \end{equation}

Analogously we call $B^-_\mu$, $m\in (-m^*,m_\beta)$ a non constant solution which is monotonic non increasing and denote by $b^-_\mu$ its asymptotic value. The existence $B_\mu$ implies the existence of $B^-_\mu$, in fact by simmetry $B^-_\mu=-B_{-\mu}$. Thus what we will say for $B_\mu$ extends to $B^-_\mu$ and in the sequel we will consider only $B_\mu$.

 \medskip
As mentioned above the existence of bumps is an open problem, the simulations indicate that bumps do indeed exist.
The relation between bump and instanton can be understood in the following way.  Call
$\bar x (\mu)$ the value of $r$ such that $\bar m(r)=\mu$.  Replace the boundary condition
$m(r)=\mu$, $r<0$, in the definition of the bump by $m(r)= \bar m(r+\bar x (\mu))$, $r<0$.  Then
the solution of  \eqref{j1.16} would be $m(r)= \bar m(r +\bar x (\mu))$, $r>0$, with $h=0$, the asymptotic value at $r=+\infty$ being $m_\beta$.  Replacing $m(r)=\mu$ by $m(r)= \bar m(r+\bar x (\mu))$ for $r<0$ is a small error if $\mu$ is close to $-m_\beta$ (because the instanton converges exponentially to its asymptotic values).  One may then hope to prove in such a case the existence of the bump using perturbative techniques as in
\cite{DOP}--\cite{DOP00}.  This has been done successfully
 in \cite{DOP00} for the equation
 \begin{equation*}
   m(r) = \tanh\{ \beta[ J^{\rm neum}*m(r)+h]\},\quad r \ge 0
 \end{equation*}
where $J^{\rm neum}$ is defined with Neumann conditions; $h$ above is fixed and sufficiently small.

\begin{figure}[h]
\centering
\includegraphics[width=0.70 \textwidth]{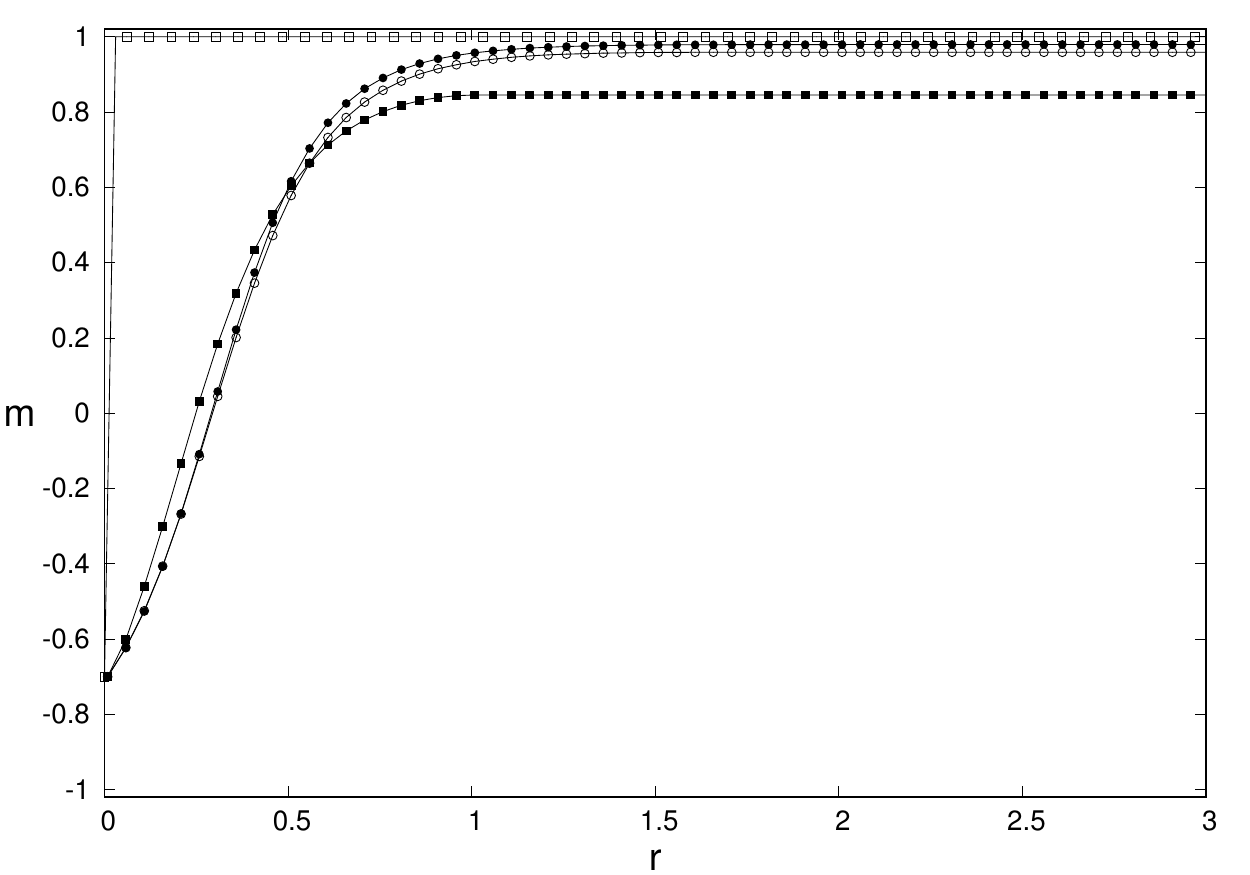}
\caption{{\footnotesize  Iterations of Eq. \eqref{j3.11}, with $\beta=2.5$, $\gamma^{-1}=120$, $\ell=5$ and $\mu=-0.7$. The different points denote, respectively, the initial condition (empty squares), the first iteration (black  squares), the second iteration (empty circles) and the third iteration (black  circles).}}
\label{fig:j4.1}
\end{figure}

We have numerical evidence of the existence of bumps.  We have simulated  \eqref{j3.11} by looking at its discrete version with $\ga^{-1}=120$, $\ell=5$ and Neumann conditions at the right boundary.  We have solved such an equation by iteration:
we start  with $m\equiv 1$, compute  $h$ via \eqref{j3.11bis}  with such $m$ and then define the first iterate
$m_1$ as $m_1= \tanh{\beta(J*m+h)}$.  We then repeat the procedure till we find a  fixed point.  This is indeed reached (approximately) after a few iterations (in fact, three iterations already suffice to obtain good numerical convergence), see Fig. \ref{fig:j4.1}.

\begin{figure}[h]
\centering
\includegraphics[width=0.49 \textwidth]{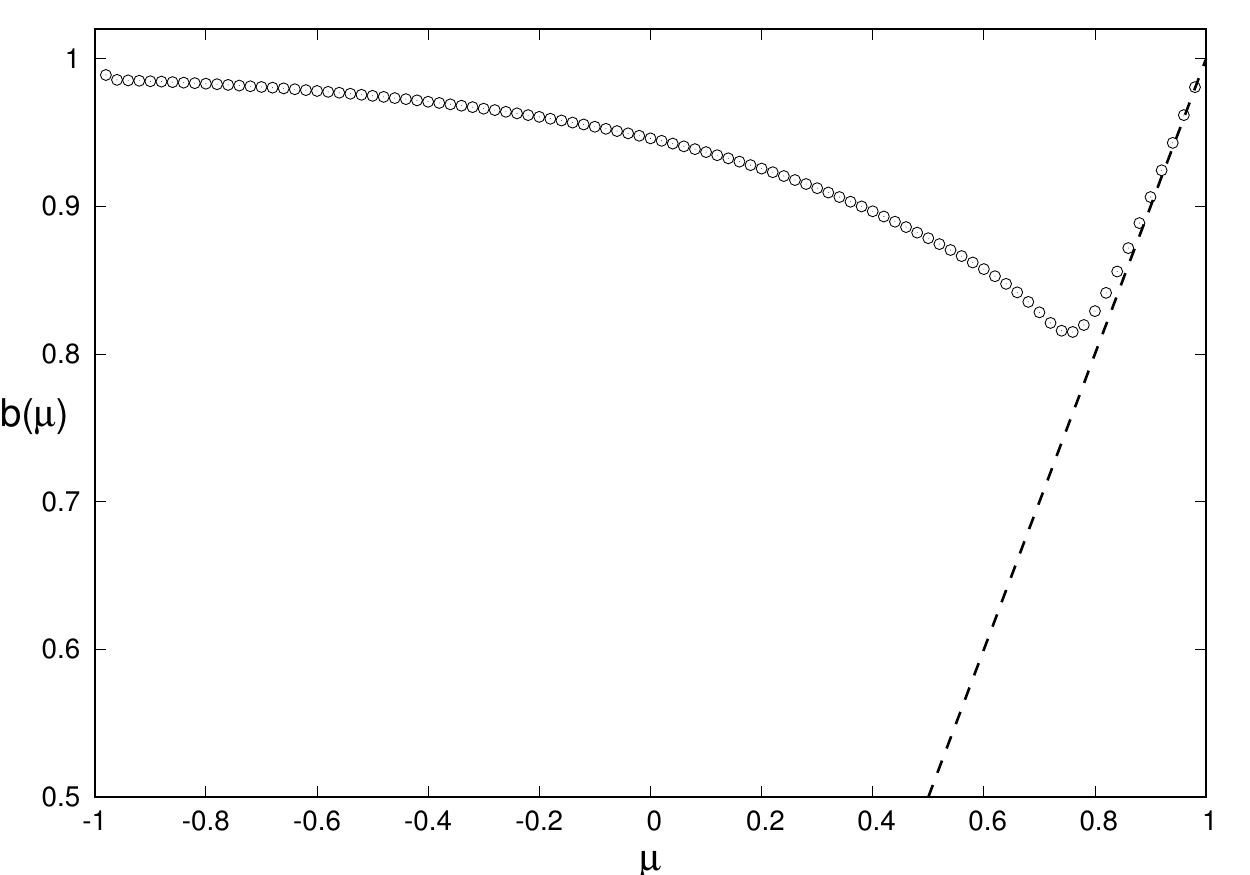}
\hskip 3 pt
\includegraphics[width=0.49 \textwidth]{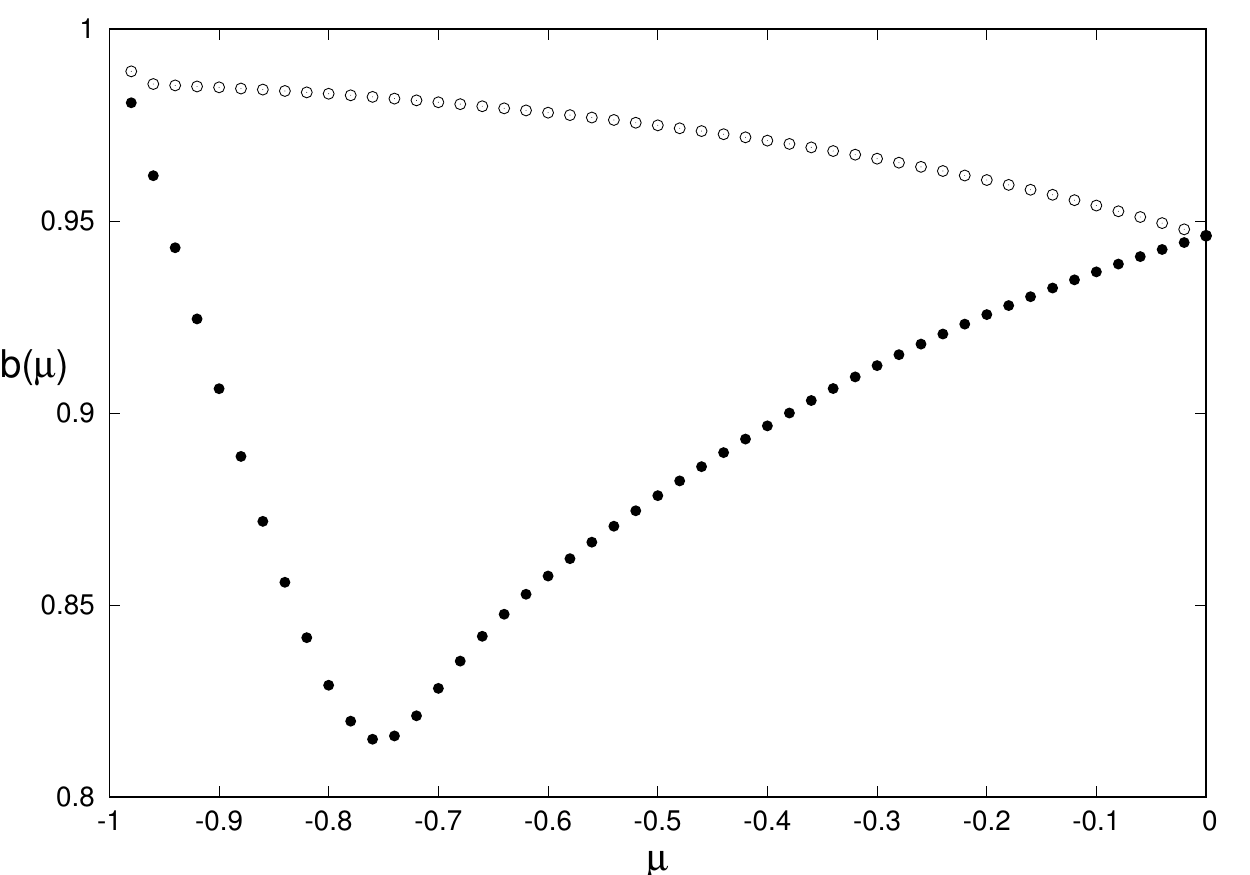}
\caption{{\footnotesize \textit{Left panel}: Behavior of $b(\mu)$, with $\beta=2.5$ and $\gamma^{-1}=120$. The black dashed line denotes the curve $b(\mu)=\mu$.
\textit{Right panel}: For any $\mu \in (-m_\beta,0)$ we report with an empty circle the value of  $b (m_-)$ and with  a black circles the value of $b(m_+)$, $m_+=-m_-$.}}
\label{fig:j4.2-3}
\end{figure}

The numerical values of $b(\mu)$ are reported in Fig. \ref{fig:j4.2-3}, the main features are:
\begin{itemize}
\item the values of  $b(\mu)$ are all in the metastable region,
\item $b(m_+)=m_+$ if $m_+\in(m^*,m_\beta)$, i.e. in the plus metastable region (left panel)
\item $b(m_-)>b(m_+)$ for $m_-\in(-m_\beta, 0)$ and $m_+=-m_-$ (right panel).
	\end{itemize}

\medskip

{\bf Conjecture}  {\em The bump $B_\mu$ exists for all $\mu  < m^*$, when $\mu \in [m^*,m_\beta)$ there is no bump and we call $b(\mu)=\mu$.
When $m_+ < m_\beta$ for all $\ell$ large enough there is a stationary solution  $m_{\rm st}(r;\ell;m_{\pm})$
of \eqref{j1.16}, such that
 \begin{equation}
\label{j3.13}
\lim_{\ell \to \infty}m_{\rm st}(r\ell;\ell;m_{\pm})=m_{\rm st}(r;m_{\pm}),\quad r\in (0,1)
\end{equation}
 \begin{equation}
\label{j3.14}
m_{\rm st}(0;m_{\pm})= b(m_-),\quad m_{\rm st}(1;m_{\pm})= b(m_+)
\end{equation}
and}
 \begin{equation}
		\label{j3.15}
-\frac 12 [1-\beta(1-m_{\rm st}^2)]\frac{dm_{\rm st}}{dr} =I_{\rm st}(m_+),\quad r\in[0,1]
\end{equation}

\medskip

{\bf Remark.}
{\em Under the above Conjecture  the channel has a positive current if}
  \begin{equation}
\label{j3.16}
b(\mu)>-\mu, \;  \mu \in (-m_\beta,-m^*);\quad
b(\mu)>b(-\mu) , \;  \mu \in (-m^*,0)
\end{equation}

 \medskip
As shown in  the right panel of Fig. \ref{fig:j4.2-3} there is clear numerical evidence of the validity of \eqref{j3.16}.\\

\begin{figure}[h]
\centering
\includegraphics[width=0.70\textwidth]{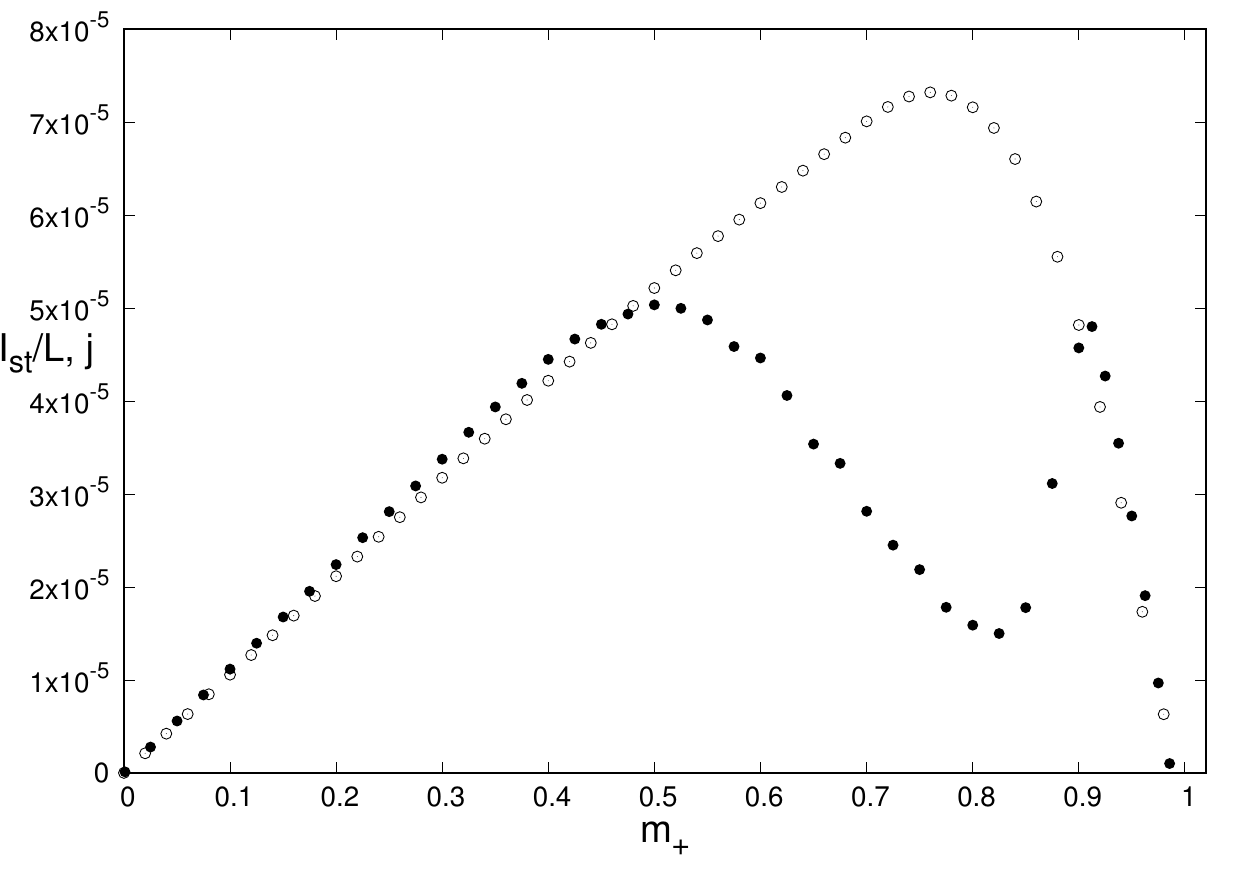}
\caption{{\footnotesize We plot $I_{\rm st}/L$  (empty circles) and $j:=j^{T}_{{\rm ch} \to\mathcal R_2}$ (black circles) as functions of $m_+$ .}}
\label{fig:10b}
\end{figure}

The equation \eqref{j3.15} with boundary conditions \eqref{j3.14}  can be  easily solved analytically thus determining $ I_{\rm st}(m_+)$.  By using the numerical values obtained for $b(m_-)$ and $b(m_+)$ we get the graph shown with empty circles in Fig. \ref{fig:10b}, where however $ I_{\rm st}$ is divided by $L$ in order to compare it with the experimental value $j(m_+)$ (black circle in Fig. \ref{fig:10b}) as given in  Fig. \ref{fig:2}. The agreement is good except in the interval $m_+\in (m',m''')$,
such a discrepancy will be discussed in the next section.

\setcounter{equation}{0}

\section{Stability of the bump}
\label{sec.j9}

The numerical analysis of \eqref{j3.11} suggests the following: 

\begin{itemize}

\item there exists a bump solution $B_\mu$ for all $\mu\in(-m_\beta,m^*)$,

\item  when $\mu\in (-m_\beta,-m^*)$ there are two solutions: $m(x)\equiv \mu$ and
$m(x)=B_\mu(x)$,

\item when
$\mu\in (-m^*,0)$ there are three solutions: $m(x)\equiv \mu$,
$m(x)=B_\mu(x)$ and $m(x)=-B_{-\mu}(x)$,

\end{itemize}

\begin{figure}[h]
\centering
\includegraphics[width=0.70\textwidth]{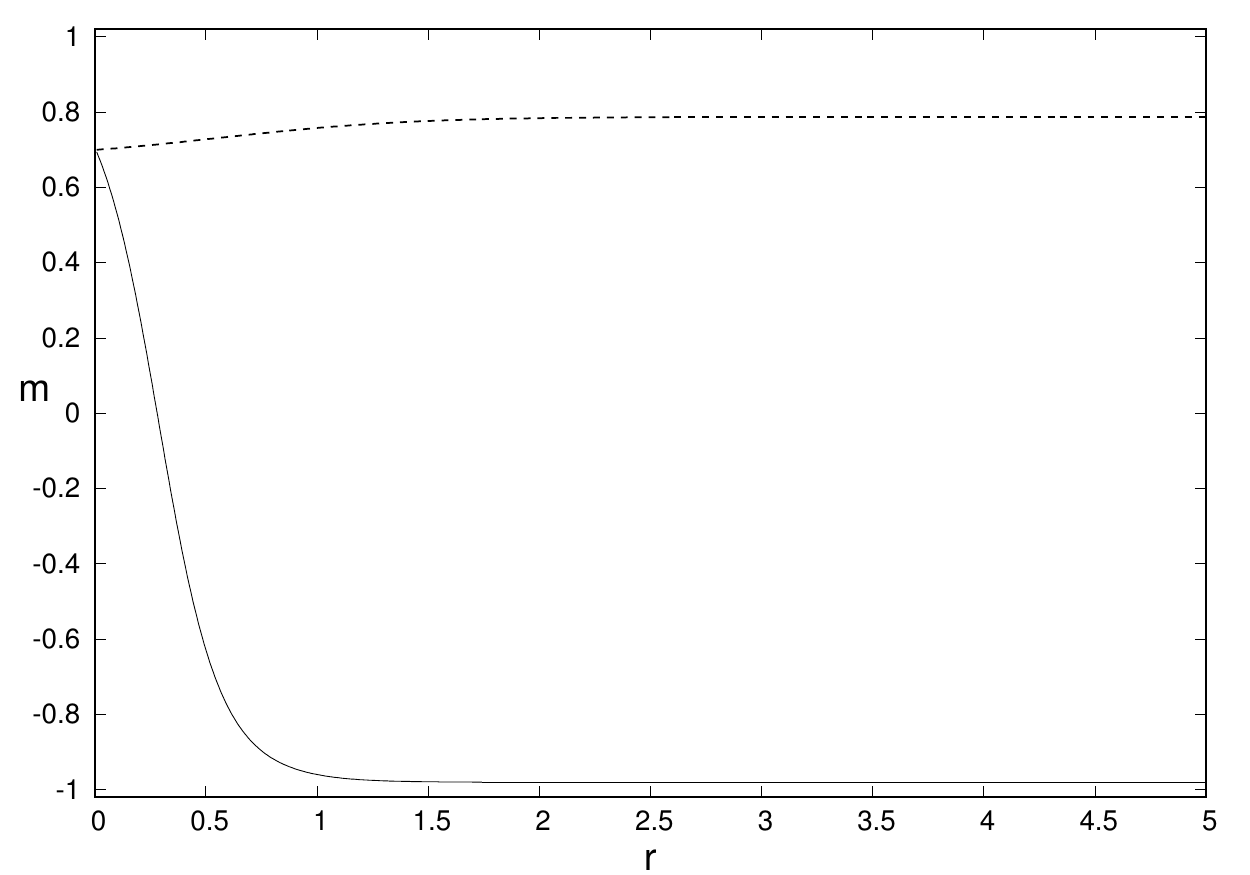}
\caption{{\footnotesize
$\beta=2.5$, $\gamma^{-1}=120$, $\ell=5$ and $\mu=0.7\in (m^{\text{iv}},m'')$. The  black dashed line represents $B_\mu$, 
the black thin line represents the asymptotic pattern  of the CA which is close to $-B_{-\mu}$. }}
\label{fig:j4.4}
\end{figure}

\begin{figure}[h]
\centering
\includegraphics[width=0.49 \textwidth]{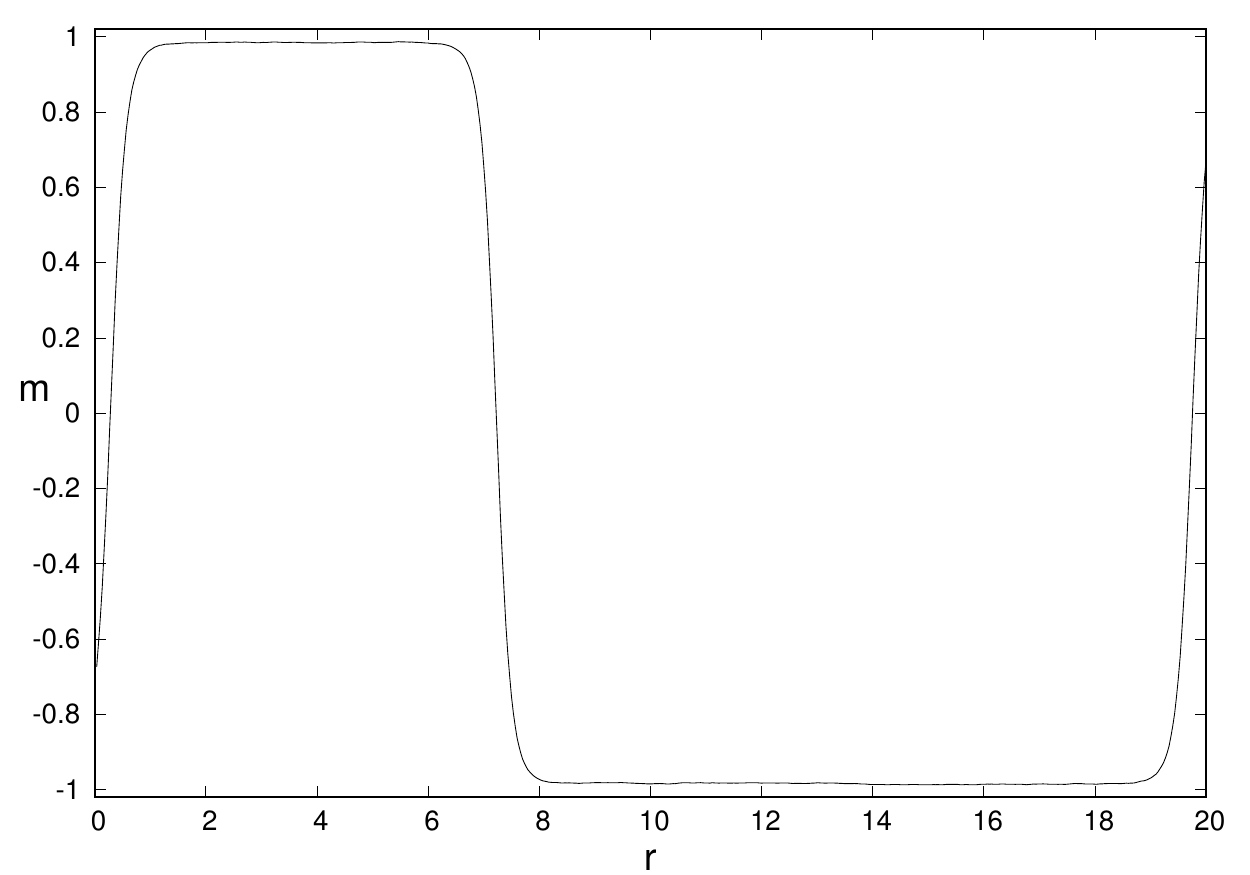}
\hskip 3 pt
\includegraphics[width=0.49 \textwidth]{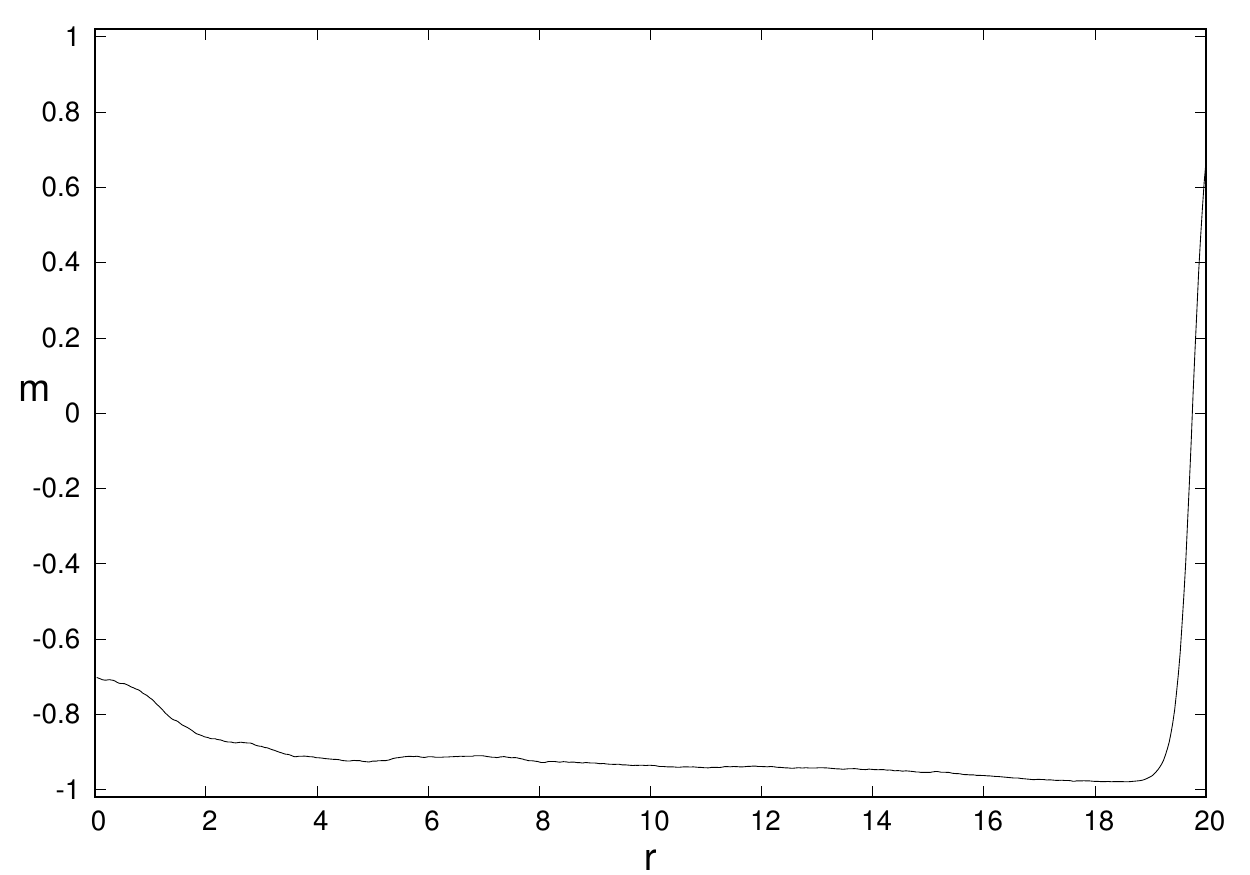}
\caption{{\footnotesize Typical (average) magnetization profile obtained at the beginning of a cycle (left panel) and at the end of a cycle (right panel).}}
\label{fig:j4.7-8}
\end{figure}

An alternative way to study the existence of the bump is by
running the OS-CA with boundary conditions $\mu$ on the left and Neumann  on the right.  We take the same parameters $\ga^{-1}=120$ and $\ell=5$
 used for the numerical analysis of the solutions of \eqref{j3.11} and start with an initial condition where all sites in the channel are occupied. Referring to Fig. \ref{fig:2}, when $\mu\in (-m_\beta,m')$ 
we see, after a transient, a steady pattern close to $B_\mu$. When $\mu\in (m^{\text{iv}},m_\beta)$ we see, after a transient, a steady pattern close to $m(x)\equiv \mu$.
When $\mu\in (m',m^{\text{iv}})$
the final pattern is close to  $-B_{-\mu}$, see Fig. \ref{fig:j4.4}.
Observe that the OS-CA does not select the bump solution when $\mu\in (m',m^{\text{iv}})$ which is approximately the region where there is discrepancy between the theoretical and the experimental curves in Fig \ref{fig:10b}. We conjecture that this is due to $\ga$ being not small enough so we are far from the mesoscopic regime and stochastic fluctuations are relevant. Stochastic fluctuations may then determine  tunnelling from the bump  to patterns where there is a bump on the left and a minus bump on the right with an instanton in between them
and patterns where the two bumps are both up.  Indeed we have numerical evidence of all that, in the times of the simulations we see in fact the magnetization patterns oscillate as described above, see Fig.\ref{fig:j4.7-8}.\\

\section{Conclusions}
\label{sec:concl}

We have presented
two sets of simulations: the first one, see  Fig. \ref{fig:6}, shows that in the CC-CA there is a non zero current $j^{CC}(\ga p)$ (provided the rate $\ga p$ of exchanges between reservoirs is in some non zero interval); the second set of simulations, see Fig. \ref{fig:2}, refers to the OS-CA at magnetization $m_+ = -m_->0$ and shows that  when $m_+ \in (0,m_\beta)$ then the current $j(m_+)$ goes ``in the wrong direction'', namely from the reservoir with $m_-$ to that with $m_+$.  We have a heuristic proof that what seen in  Fig. \ref{fig:6} follows from the behavior of the channel in the OS-CA, as shown in Fig. \ref{fig:2}; the proof relies on the validity of the mesoscopic and the adiabatic limits.

In the case of  Fig. \ref{fig:2} the current is negative when $m_+>m_\beta$ and positive when $m_+<m_\beta$, in the former case the magnetization pattern in the channel shows the coexistence of the plus and minus phases while in the latter case only one phase appears (the statement in both cases refers to what happens in most of the volume).
When the current is negative the values of the magnetization in the plus phase are larger than $m_\beta$ and smaller than $-m_\beta$ in the negative one.  Instead when the current is positive we are in the one phase regime and the values of the magnetization are metastable (thus there is a state with positive current which in the bulk takes positive metastable values and another state also  with positive current which in the bulk takes negative metastable values).
When the current is negative the plus and minus phases are connected via an instanton-like
profile around the center of the channel, when the current is negative the unstable values of the magnetization are localized  in a small region close to the endpoints.  We thus have a boundary layer which leads quite abruptly from the imposed values of the magnetization at the boundaries to some metastable value after which the magnetization pattern is smooth and the current flows opposite to the magnetization gradient in agreement with the Fourier law.

The strange phenomenon of the current going  ``in the wrong direction'' depends on the fact that the magnetization-jump in the boundary layer is more pronounced if it starts from lower values of the magnetization.  Such a property, see \eqref{j3.16}, follows from the solution (the bump) of a non local equation describing the boundary layer, but its solution is obtained only numerically and we do not have a mathematical proof or even a heuristic explanation of why \eqref{j3.16} should hold.

We expect that also in the Cahn-Hilliard equation the graph of $j(m_+)$ has a qualitatively similar shape as in  Fig. \ref{fig:2}, but we miss a proof.

We imagine that our results extend to more general systems with Kac potentials and maybe to physical systems where a van der Waals type of phase transition is present.  In such cases a metastable interval is well defined and the relevant density (or magnetization) patterns  in the bulk of the channel should have metastable values.  Also for short range interactions, as in the n.n.\ Ising model with ferromagnetic interactions there are metastable values but the metastable region depends on the size of the system and shrinks to 0 as the volume diverges.  Take the 2D Ising model in a squared box of side $L$: in the periodic case for $\beta$ large it is proved that if $\pm m_\beta$ are the equilibrium magnetizations then for
the canonical Gibbs measure with average magnetization $m \in (-m_\beta, -m_\beta + c L^{-2/3})$ and $c$ small enough the phenomenon of phase separation is absent.  Consider the Kawasaki dynamics at such values of $\beta$ with periodic conditions on the horizontal sides of the box
and exchanges of the spins in the vertical ones with infinite reservoirs at magnetization $m_-$ and $m_+$ on the left and right.  If what we have observed extends to this 2D Ising model we should see in the bulk magnetization patterns in the metastable phase, hence with values in an interval of size $L^{-2/3}$.  The current should therefore scale as  $L^{-1}L^{-2/3}= L^{-5/3}$ and if the boundary layer goes like in our case then the current would go from the small to the large values of the reservoirs magnetization.

\appendix
\setcounter{equation}{0}

\section{Estimates on the current between reservoirs}
\label{aappA}

Recalling \eqref{2.9ab} we have
		\begin{equation}
		\label{C.1}
  j_{\mathcal R_2\to \mathcal R_1}(t)
=j_t := \zeta_t \sum_{i_+,i_-}\mathbf 1_{\xi_t=(i_+,i_-)}[\theta''_t(i_+)- \theta''_t(i_-)]
		\end{equation}
where $ \zeta_t$ and $\xi_t$ are random variables independent of the process till time $t$ and of $\theta''_t$, they are also independent of each other. $ \zeta_t$ takes value 1 with probability $\ga p$ and value 0 with probability $1-\ga p$; the values of $\xi_t$ are pairs
$(i_+,i_-)$, $i_+\in \mathcal R_2$, $i_-\in \mathcal R_1$ and $P(\xi_t=(i_+,i_-))=\frac 1{R^2}$. The sum $ \sum_{i_+,i_-}$ is over $i_+\in \mathcal R_2$ and $i_-\in \mathcal R_1$.  We first estimate the expected value of $ j_{\mathcal R_2\to \mathcal R_1}(t)$:
	\begin{equation}
		\label{C.1a}
 E_\ga[ j_{\mathcal R_2\to \mathcal R_1}(t)]=
E_\ga[\frac{N''_{\mathcal R_2}(t)-N''_{\mathcal R_1}(t)}{R}]\ga p
		\end{equation}
where
	\begin{equation}
		\label{C.1b}
N''_{\mathcal R_i}(t)=\sum_{i\in\mathcal R_i}\theta''_t(i),\quad i=1,2
		\end{equation}
Since $|N''_{\mathcal R_i}(t)-N_{\mathcal R_i}(t)|\le 2$ for all $t$ we have
	\begin{equation}
		\label{C.1c}
\Big| E_\ga[ j_{\mathcal R_2\to \mathcal R_1}(t)]-\ga p
E_\ga[\frac{N_{\mathcal R_2}(t)-N_{\mathcal R_1}(t)}{R}]	\Big|\le \ga p\frac 4R	\end{equation}

We will next prove \eqref{C.0}.
Since $\theta''$ has values 0, 1 we have from \eqref{C.1}
\begin{equation}
\label{C.2}
E[ j_t ]\le \ga p
\end{equation}

By \eqref{j1.20.3.1} the left hand site of \eqref{C.0},  can be written as
\begin{equation}
\label{C.3}
A_T:=E\Big[\{ \frac 1T \sum_{t=0}^{T-1} [j_t- \ga p R^{-1}(N_{+,t}-N_{-,t}) ] \}^2\Big]
\end{equation}
where
\begin{equation}
\label{C.4}
N_{+,t}= \sum_{i_+}\eta_t(i_+)=N_{\mathcal R_2}(t),
\quad N_{-,t}= \sum_{i_-}\eta_t(i_-)=N_{\mathcal R_1} (t)
\end{equation}
Define $N''_{\pm,t}$ as in \eqref{C.4} but with $\theta''_t$ instead of $\eta_t$ and $A''_T$
as in \eqref{C.3} but with  $N''_{\pm,t}$.

\medskip

\begin{lem}
\begin{equation}
\label{C.5}
A_{T} \le A''_T + \frac {16} R (\ga p)^2 + \frac{16}{R^2} (\ga p)^2
\end{equation}

\end{lem}

\medskip
\noindent
{\bf Proof.}  Call
\begin{equation}
\label{C.6}
a_{t} =j_t - \ga p R^{-1}(N''_{+,t}-N''_{-,t})
\end{equation}
\begin{equation}
\label{C.7}
b_{t} =  \ga p R^{-1}\{(N''_{+,t}-N''_{-,t}) - (N_{+,t}-N_{-,t})\}
\end{equation}
Then
\begin{equation}
\label{C.8}
A_{T} =  E\Big[\frac 1{T^2} \sum_{s,t} (a_{t}-b_{t})(a_{s}-b_{s})\Big]
\end{equation}
Hence
\begin{equation}
\label{C.9}
A_{T} \le A''_{T} +2 E\Big[\frac 1{T^2} \sum_{s,t}|a_{t}||b_{s}|\Big]
+ E\Big[\frac 1{T^2} \sum_{s,t}|b_{s}| |b_t|\Big]
\end{equation}
$|b_t|\le \ga p \frac 4R$ because $|N''_{+,t} - N''_{-,t}| \le R$ and $|N''_{\pm,t} - N_{\pm,t}| \le 2$.  By \eqref{C.2} and $|N''_{+,t} - N''_{-,t}| \le R$ we get $E_\ga[|a_t|] \le 2 \ga p$, therefore
\begin{equation}
\label{C.10}
A_{T} \le A''_{T} +2 \ga p \frac 8 R \ga p  + [\ga p \frac 4R]^2
\end{equation}
 \qed

\medskip

\begin{lem}
Let $s<t$ and $a_t$ as in \eqref{C.6} then
\begin{equation}
\label{C.11}
E\Big[ a_s a_t\Big] =0
\end{equation}

\end{lem}

\medskip
\noindent
{\bf Proof.}  By the independence properties of $\zeta_t$ and $\xi_t$:
\begin{equation}
\label{C.12}
E [ a_s j_t ] = E \Big[ a_s \ga p \sum_{i_+,i_-} R^{-2}[\theta''_t(i_+)- \theta''_t(i_-)]\Big] =  E \Big[ a_s \ga p R^{-1}[N''_{+,t}- N''_{-,t}]\Big]
\end{equation}
\qed

\medskip
As a consequence
\begin{equation}
\label{C.13}
 A''_T = \frac 1{T^2} \sum_{t=0}^{T-1} E[a_t^2
]
\end{equation}
We expand the square in $E[a_t^2
]$, the first term is
\begin{equation*}
 E\Big[ \zeta_t \sum_{i_+,i_-}\sum_{i'_+,i'_-}\mathbf 1_{\xi_t=(i_+,i_-)}
 \mathbf 1_{\xi_t=(i'_+,i'_-)}[\theta''_t(i_+)- \theta''_t(i_-)][\theta''_t(i'_+)- \theta''_t(i'_-)]\Big]
\end{equation*}
Due to the characteristic functions $i_{\pm}=i'_{\pm}$ so that the above is bounded by
$\ga p$.  The double product in the expansion of $E[a_t^2
]$ is bounded by $2(\ga p)^2$ and the third term by $(\ga p)^2$, so that
\begin{equation}
\label{C.14}
 A''_T \le  \frac 1T \{\ga p + 3(\ga p)^2\}
\end{equation}
Going back to \eqref{C.10} we get
\begin{equation}
\label{C.15}
A_{T} \le \frac 1T \{\ga p + 3(\ga p)^2\} +16  \frac {(\ga p)^2}R  + 16[ \frac{\ga p}R]^2
\end{equation}
which concludes the proof of \eqref{C.0}.

\setcounter{equation}{0}
\section{Proof of Theorems \ref{thm5.1} and \ref{thm5.1b}}
\label{A1}

{\bf Proof of \eqref{j1.16}. } Here we prove that $m(r,t)$ satisfies \eqref{j1.16} both in the CC-CA and in the OS-CA.

\medskip
Let $u(r,t)=m(r,t)+1$ then  $m$ satisfies  \eqref{j1.16} if and only if $u$ satisfies
	\begin{eqnarray}
		\label{B.1a}
 &&\frac{\partial}{\partial t}u(r,t) =  \frac 12
 \frac{\partial^2 u}{\partial r^2}  - C \frac{\partial }{\partial r} \Big\{ [u(2-
 u]\int_{r}^{r+1}[u(r+\xi,t)
 - u(r-\xi,t)] d\xi\Big\}
 \end{eqnarray}
with $u(r+\xi,t)= u_+(t)=m_+(t)+1$ if $r+\xi\ge \ell$ and $u(r-\xi,t)= u_-(t)=m_-(t)+1$ if $r-\xi\le 0$.  In the OS-CA $m_\pm(t)\equiv m_\pm$.

By \eqref{5.3a}
	\begin{eqnarray}
	\nn
	&& \lim_{\ga \to 0} \lim_{\ga x\to r, \ga^{2}t\to \tau}  E_\ga[ \eta(x,v,t)] =\frac 12 u(x,t), \qquad v\in\{-1,1\}
\\&& \lim_{\ga \to 0} \lim_{\ga x\to r, \ga^{2}t\to \tau}  u_\ga(x,t)=u(x,t),\qquad   u_\ga(x,t)=E_\ga[ \eta(x,t)] \\\label{B.1b}
\nn
	\end{eqnarray}
So that we need to prove that the limit of $u_\ga$ satisfies \eqref{B.1a}.

By assumpotion $u(r,t)$ is smooth so that it is enough to prove weak convergence namely that
for any smooth test function $f(r,t)$ with compact support in $(0,\ell)\times (0,\infty)$,
\begin{eqnarray}
\label{B.2b}
&&\hskip-1cm \int u(r,t)\frac{\partial  f(r,t)}{\partial t} dr dt=  - \frac 12 \int u(r,t)\frac{\partial^2 f(r,t)}{\partial r^2}  dr dt \nn\\&&\hskip.1cm - \int
 \frac{\partial f(r,t)}{\partial r} C \Big\{ [u(2-
 u]\int_{r}^{r+1}[u(r+\xi,t)
 - u(r-\xi,t)] \Big\}dr dt
\end{eqnarray}
By an integration by parts
		\begin{eqnarray*}
 &&\int u(r,t)\frac{\partial  f(r,t)}{\partial t} dr dt=-\lim_{\ga \to 0} \ga^3\sum_{x,t} f(\ga x,\ga^2t)
 \ga^{-2}\{u_\ga(x;t+1)- u_\ga(x;t)\}
	\end{eqnarray*}
We will next consider $u_\ga(x;t+1)- u_\ga(x;t)$. Recalling that	$j_{x,x+1}(t)$ is the number of particles which in the time step $t, t+1$  cross the bond $(x,x+1)$, $x\in\{1,..,L-1\}$ (counting as positive those which jump from $x$ to $x+1$ and as negative those from $x+1$ to $x$), we have
	\begin{equation*}
	u_\ga(x;t+1)- u_\ga(x;t)=E_\ga[j_{x-1,x}(t)]-E_\ga[j_{x,x+1}(t)]
	\end{equation*}
We then have denoting by  $\nabla_\ga$ the discrete derivative ($\nabla_\ga \varphi(x)=\varphi(x+1)-\varphi(x)$),
	\begin{equation}
	\label{B.5}
\int u(r,t)\frac{\partial  f(r,t)}{\partial t} dr dt=-\lim_{\ga \to 0} \ga^3\sum_{x,t}  \ga^{-1} \nabla_\ga f(\ga x,\ga^2t)
 \ga^{-1}E_\ga[j_{x,x+1}(t)]
	\end{equation}
\begin{lem}
\label{lemmaA3}
	\begin{equation}
\label{B.4}
E_\ga[j_{x,x+1}(t)]= \frac 12[u_\ga(x;t)- u_\ga(x+1;t)+E_\ga\Big[\chi_{x,\ga;t}
\eps_{x,\ga;t}+\chi_{x+1,\ga;s}
\eps_{x+1,\ga;t}\Big]
\end{equation}
where $\eps_{x,\ga;t}$ is  $\eps_{x,\ga}$ computed at time $t$ and
\[
\chi_{x,\ga;t} = \eta (x,1;t)\Big(1- \eta (x,-1;t)\Big)+
 \eta (x,-1;t)\Big(1- \eta (x,1;t)\Big)
\]

\end{lem}

\medskip
\noindent
{\bf Proof.}  Observe that the expected number of particles that goes from $x$ to $x+1$  is
		\begin{equation*}
E_\ga\Big[\eta(x,1;t)\eta(x,-1,t)+\chi_{x,\ga;t}(\frac 12+\eps_{x,\ga;t})\Big]=\frac 12 u_\ga(x,t)+ E_\ga\Big[\chi_{x,\ga;t}\eps_{x,\ga;t}\Big]
	\end{equation*}
The expected number of particles that goes from $x+1$ to $x$ is
		\begin{eqnarray*}
&&E_\ga\big[\eta(x+1,1;t)\eta(x+1,-1,t)+\chi_{x+1,\ga;t}(\frac 12-\eps_{x+1,\ga;t})\big]=\frac 12 u_\ga(x+1,t)
\\&& \hskip7cm - E_\ga\big[\chi_{x+1,\ga;t}\eps_{x+1,\ga;t}\big]
	\end{eqnarray*}
 so that we get \eqref{B.4}.\qed

 \medskip
We insert \eqref{B.4} in \eqref{B.5} and, denoting by  $\Delta_\ga$ the discrete laplacian, we get
	\begin{eqnarray}
	\nn
&&
\ga^3\sum_{x,t}  \ga^{-1} \nabla_\ga f(\ga x,\ga^2t)
 \ga^{-1}j_\ga(x,x+1,t)= \frac 12 \ga^3\sum_{x,t}\ga^{-2} \Delta_\ga f(\ga x,\ga^2t)
u_\ga(x,t)\\&&\hskip1cm + \ga^3\sum_{x,t} \ga^{-1} 2 f'(\ga x,\ga^2t)E_\ga[\chi_{x,\ga;t}]
E_\ga[\ga^{-1}\eps_{x,\ga;t}] +R_\ga \nn\\
\label{B.6}
	\end{eqnarray}
where $2 f'(\ga x,\ga^2t)= [\nabla_\ga f(\ga x,\ga^2t)+ \nabla_\ga f(\ga (x-1),\ga^2t)]$ and
	\begin{eqnarray*}
R_\ga :=2 \ga^3\sum_{x,t} \ga^{-1} 2 f'(\ga x,\ga^2t) E_\ga[\chi_{x,\ga;t}\Big(\ga^{-1}\eps_{x,\ga;t}-
E_\ga[\ga^{-1}\eps_{x,\ga;t}]\Big)]
	\end{eqnarray*}
By \eqref{B.1b} and \eqref{j1.15.3}
	\begin{eqnarray}
	\nn&& \lim_{\ga \to 0} \lim_{\ga x\to r, \ga^{2}t\to \tau}  E_\ga[\chi_{x,\ga;t}]=\frac 12 u(r,t)[2-u(r,t)]\\ \label{B.8}
\\&& \lim_{\ga \to 0} \lim_{\ga x\to r, \ga^{2}t\to \tau}  E_\ga[\ga^{-1}\eps_{x,\ga;t}] = \int_{r}^{r+1}  C[u(r+\xi,t)
 - u(r-\xi,t)]d\xi \nn
	\end{eqnarray}
We postpone the proof of 
	\begin{equation}
\label{A1.6}
 \lim_{\ga \to 0} \ga^3\sum_{x=2}^{\ga^{-1}\ell-1} \sum_{t=1}^{\ga^{-2}T}  E_\ga\Big[\big|\ga^{-1}\eps_{x,\ga;t}-
E_\ga[\ga^{-1}\eps_{x,\ga;t}]\big| \Big] = 0
\end{equation}
where  $(0,\ell)\times (0,T)$ contains the support of $f(r,t)$. 

Observe that \eqref{B.5}, \eqref{B.6}, \eqref{B.8} and \eqref{A1.6} yield \eqref{B.2b} concluding the proof of \qed

\medskip
{\bf Proof of \eqref{A1.6}.} By Cauchy-Schwartz it is enough to prove that
\begin{equation}
\label{A1.6.1}
 \lim_{\ga \to 0} \ga^3\sum_{x=2}^{\ga^{-1}\ell-1} \sum_{t=1}^{\ga^{-2}T}  E_\ga\Big[\big |\ga^{-1}\eps_{x,\ga;t}-
E_\ga[\ga^{-1}\eps_{x,\ga;t}]\big|^2 \Big] = 0
\end{equation}
We thus need to compute the limit of
\begin{equation}
\label{A1.6.2}
\ga^5\sum_{r,r',r''\in \ga \mathbb Z} \sum_{\tau\in \ga^2 \mathbb Z}
g_\ga(r,r',r'',\tau)
\end{equation}
where $\ga^{-1}r \in [2,\ga^{-1}\ell-1]$, $|r'-r| \le 1$, $|r''-r| \le 1$,
$\ga^{-2}\tau \in [1,\ga^{-2}T]$ and
\begin{equation}
\label{A1.6.3}
g_\ga(r,r',r'',\tau) = C^2E_\ga[\tilde \eta_{\ga^{-2}\tau} (\ga^{-1} (r'-r))
\tilde \eta_{\ga^{-2}\tau} (\ga^{-1} (r''-r))   ]
\end{equation}
where $\tilde \eta_t(x) = \eta (x,t)-E_\ga[\eta (x,t)]$ if $x\in [1,L]$, otherwise
it is  $=\frac {2 N_{\mathcal R_i}}{R}- E_\ga[\frac {2 N_{\mathcal R_i}}{R}]$ where
$i=2$ if $x>L$ and $i=1$ if $x<1$ otherwise in the OS-CA is equal to $m_\pm$ respectively.
By \eqref{j1.15.3} and \eqref{j1.15.44}, 
 \eqref{A1.6.2} vanishes  as $\ga\to 0$. \qed

\medskip

\noindent
{\bf Proof of \eqref{B.12}.}
We call
	\begin{equation}
\label{B.13}
 I_{x,\ga}^T=\ga \sum_{t=0}^{T-1} E_\ga[j_{x,x+1}(t)]
 \end{equation}
		\begin{lem}
 There are $c$ and $c'$ so that for all $r'<r''$ in $(0,\ell)$
\begin{equation}
\label{B.9.2}
\Big|\frac{1}{x''-x'}\sum_{y=x'}^{x''}I_{y,\ga}^T\Big|\le c,\qquad x'=[\ga^{-1}r'],\quad x''=[\ga^{-1}r'']
 \end{equation}
	\begin{equation}
\label{B.9.3}
\Big|I_{x'',\ga}^T-I_{x',\ga}^T\Big|\le c' |r''-r'|
	 \end{equation}
\end{lem}

\noindent{\bf Proof.} By \eqref{B.4}, using that $|\chi_{x,\ga;t}|\le 2$ and $|\eps_{x,\ga;t}| \le 2C\ga$ for all $x$ and $t$ and after telescopic cancellations
we get
	\begin{eqnarray*}
\Big|\frac{1}{x''-x'}\sum_{y=x'}^{x''}I_{y,\ga}^T\Big|\le\big| \ga \sum_{s=0}^{T-1}\frac{1}{x''-x'} E_\ga[\frac 12(\eta(x',s)-\eta(x''+1,s))]\big|+  8 C \ga^2T
	\end{eqnarray*}
The right hand side converges to $\frac 1{r''-r'}\int_0^\tau \frac 12 [m(r'.s)-m(r'',s)] ds+8C^2\tau$
which, by the smoothness of $m$, proves \eqref{B.9.2}.

We have that
	\begin{equation*}
\Big|	\ga \sum_{t=0}^{T-1} j_{x',x'+1}(t)]-\ga \sum_{t=0}^{T-1} j_{x'',x''+1}(t)\Big|\le c' \ga |x''-x'|
	\end{equation*}
because the particles which contribute to the left hand site
are: (1) those which reach for the first time $x'+1$ jumping from $x'$ and at the final time are in $[x'+1,x'']$; (2) those which reach for the first time $x''$ jumping from $x''+1$ and at the final time are in $[x'+1,x'']$;  (3) those initially in $[x'+1,x'']$ and which leave this interval for the last time jumping to $x''+1$; (4)  those initially in $[x'+1,x'']$ and which leave this interval for the last time jumping to $x'$.
\qed

\bigskip
The family $\{I_{x,\ga}^T\}$ thought as functions of $r=\ga x$ are equibounded and equicontinuous in any compact of $(0,\ell)$, thus they converge pointwise by subsequences. We will then prove \eqref{B.12} by identifying the limit. By continuity it will be enough to prove
\begin{equation}
\label{A.9.2}
\lim_{\ga\to 0}\frac{1}{x''-x'}\sum_{y=x'}^{x''} I_{x,\ga}^T=\frac{1}{r''-r'} \int_{r'}^{r''}dr\int_0^\tau I(r,s)ds
 \end{equation}
 By \eqref{B.4}
 	\begin{eqnarray*}
\frac{1}{x''-x'}\sum_{y=x'}^{x''} I_{x,\ga}^T&=&
\ga \sum_{s=0}^{T-1}\Big\{\frac{1}{x''-x'} E_\ga[\frac 12(\eta(x',s)-\eta(x''+1,s))]
\\&+& \frac{1}{x''-x'}\sum_{y=x'}^{x''}E_\ga[\chi_{x,\ga;s}\eps_{x,\ga;s}+\chi_{x+1,\ga;s}\eps_{x+1,\ga;s}]\Big\}
 	\end{eqnarray*}
 The first term converges to
\begin{equation}
\label{A.9.4}
\frac{1}{r''-r'} \frac 12\int_0^\tau [m(r',s)-m(r'',s)]ds= \frac{1}{r''-r'} \frac 12\int_0^\tau ds\int_{r'}^{r''}dr\frac{\partial m(r,s)}{\partial s}
 \end{equation}
By \eqref{B.8} and \eqref{A1.6} the second one  converges to
	\begin{equation*}
	-\frac{1}{r''-r'} \int_0^\tau \int_{r'}^{r''} C   [1-
 m^2]\int_{r}^{r+1}[m(r+\xi,s)
 - m(r-\xi,s)] d\xi dr  ds
	\end{equation*}
	
	\noindent
{\bf Proof of \eqref{j1.19.1}.} As the two are similar, we just prove the second equality in  \eqref{j1.19.1}.   The same proof as the one for \eqref{B.9.3} shows that
\begin{equation}
\label{B.16}
\Big|I_{x,\ga}^T-I_{{\rm ch}\to \mathcal R_2,\ga}^T\Big|\le c' |\ell-r|,\qquad x=[\ga^{-1}r]
	 \end{equation}
where
	\begin{equation*}
	I_{{\rm ch}\to \mathcal R_2,\ga}^T= \ga \sum_{t=0}^{T-1} E_\ga[j_{{\rm ch}\to \mathcal R_2}(t)]
		 \end{equation*}
Let $\tilde I$ be a limit point of $I_{{\rm ch}\to \mathcal R_2,\ga}^T$ as $\ga\to 0$ then
	\begin{equation*}
\Big|\int_0^\tau I(r,s)ds -\tilde I\Big|\le c' |\ell-r|
		 \end{equation*}
Using the expression \eqref{j1.16} for $I(r,t)$ and the continuity of $m$, we get in the limit $r\to \ell$ that $\tilde I=\int_0^\tau I(\ell,s)ds$.

\noindent
{\bf Proof of \eqref{j1.17}}. As the proofs are similar, we just prove the second equality in \eqref{j1.17} for the CC-CA.
Suppose by contradiction that there is $t>0$ such
that $m(\ell,t)\ne m_{+}(t)$ and for the sake of definiteness $m(\ell,t)< m_{+}(t)$.  Then
there is $\delta>0$ and an interval $[t',t'']$ so that for $s\in [t',t'']$, $m_{+}(t)> m(\ell,t) +\delta$. Recalling the proof of Lemma \ref{lemmaA3}
		\begin{eqnarray*}
	E_\ga[j_{{\rm ch}\to \mathcal R_2}(s)]&=&E_\ga[ \frac{N_{\mathcal R_2}(s)}{R}]
- \frac 12 u_\ga(L,s)- E_\ga\Big[\chi_{L,\ga;s}\eps_{L,\ga;s}\Big] \\&\ge &
E_\ga[ \frac{N_{\mathcal R_2}(s)}{R}]-\frac 12 u_\ga(L,s) -c\ga
	 \end{eqnarray*}
$c$ a suitable constant, $c\ga$ bounding the term with $\eps_{x,\ga}$.
Then, recalling \eqref{j1.13}, \eqref{j1.20.3.1}, \eqref{2.5aa} and using the assumptions in Theorem \ref{thm5.1} we get
	\begin{eqnarray*}
\liminf_{\ga \to 0}\ga^2\sum_{s\in \mathbb Z \cap \ga^{-2}[t',t'']}	E_\ga[j_{{\rm ch}\to \mathcal R_2}(s)]&\ge &\frac 12 \int_{t'}^{t''}[m_+(s)- m(\ell,s)]ds\ge \frac{\delta}2 [t''-t']
	 \end{eqnarray*}
which contradicts \eqref{j1.19.1}.

\medskip

\noindent
{\bf The dynamics of the reservoirs.}  We just
prove \eqref{j1.18}. Let $\tau_0\ge 0$, $\tau>0$, $t_0=[\ga^{-2}\tau_0]$, $T=[\ga^{-2}\tau]$, then
		\begin{eqnarray*}
N_{\mathcal R_2}(t_0+T)-N_{\mathcal R_2}(t_0)=\sum_{t=t_0}^{t_0+T-1}\Big[
j_{{\rm ch}\to \mathcal R_2}(t)-j_{\mathcal R_2\to \mathcal R_1}(t)\big]
	 \end{eqnarray*}
We take the expectation and we use \eqref{C.1c} to get
	\begin{eqnarray}
	\nn
&&\Big|E_\ga[N_{\mathcal R_2}(t_0+T)-N_{\mathcal R_2}(t_0)] -\sum_{t=t_0}^{t_0+T-1}\Big[E_\ga[j_{{\rm ch}\to \mathcal R_2}(t)]-
E_\ga[\frac{N_{\mathcal R_2}(t)-N_{\mathcal R_1}(t)}{R}]\ga p
\Big]\Big|
\\&&\hskip3cm\le \frac{4\ga p}RT\label{B.18}
	 \end{eqnarray}
	 We then get
\begin{eqnarray}
	\nn
\frac a2 [m_+(\tau_0+\tau)-m_+(\tau_0)]=\int_{\tau_0}^{\tau_0+\tau} I(\ell,s)ds-p\int_{\tau_0}^{\tau_0+\tau} \frac 12[m_+(s)-m_-(s)]ds
\\ \label{B.19}
	 \end{eqnarray}
which is obtained from \eqref{B.18} by multiplying by $\ga$ and taking the limit $\ga\to 0$ after  using  that (1) $R=a\ga^{-1}$,  (2)  by \eqref{j1.15.2}
	\begin{eqnarray*}
\lim_{\ga \to 0}E_\ga[\frac{N_{\mathcal R_2}(t)-N_{\mathcal R_1}(t)}R]=\frac{m_+(\tau)-m_-(\tau)}2, \quad t	=[\ga^{-2}\tau]
	\end{eqnarray*}
(3) by \eqref{j1.19.1}
	\begin{eqnarray*}
\lim_{\ga\to 0}\ga \sum_{t=t_0}^{t_0+T-1} E_\ga[j_{{\rm ch}\to \mathcal R_2}(t)]=\int_{\tau_0}^{\tau_0+\tau} I(\ell,s)ds
	\end{eqnarray*}
Then \eqref{j1.18} is obtained from \eqref{B.19} by dividing by $\tau$ and taking the limit $\tau\to 0$.

\bigskip

\textbf{Acknowledgements} The authors acknowledge very useful discussions with Dima Ioffe.


\begin{thebibliography}{3}

\bibitem{bruno} P. Bruno, {\em Comment on "Quantum Time Crystals"}, Phys. Rev. Lett. {\bf 110}, 118901 (2013).

\bibitem{Bunim01} L. A. Bunimovich, {\em Mushrooms and other billiards with divided phase space}, Chaos \textbf{11}, 802--808 (2001).

\bibitem{CDP}  M. Colangeli, A. De Masi, E. Presutti, {\em Latent heat and the Fourier law}, Physics Letters A \textbf{380}, 1710--1713 (2016).

\bibitem{daipra} P. Dai Pra , M. Fischer, D. Regoli, {\em A Curie-Weiss Model with Dissipation}, Journal of Statistical Physics \textbf{152}, 37--53 (2013).

\bibitem{DPT}
A. De Masi, E. Presutti, D. Tsagkarogiannis, {\em  Fourier law, phase transitions and the stationary Stefan problem}, Archive for Rational Mechanics and Analysis \textbf{201}, 681--725 (2011).

\bibitem{DO} A. De Masi, S. Olla,  {\em Quasi-static hydrodynamic limit}, Journal of Statistical Physics {\bf 161}, 1037--1058 (2015).

\bibitem{DOP}
A. De Masi, E. Olivieri, E. Presutti, {\em Spectral properties of integral operators in problems of interface dynamics and
metastability}, Markov Processes and Related Fields {\bf 4}, 27--112, (1998).

\bibitem{DOP00}
A. De Masi, E. Olivieri, E. Presutti, {\em Critical droplet for a non local mean field equation},
 Markov Processes and Related Fields {\bf 6}, 439--471 (2000).

\bibitem{giacomin} G. B. Giacomin, C. Poquet, {\em Noise, interaction, nonlinear dynamics and origin of rhythmic behaviors}, Brazilian Journal of Probability and Statistics \textbf{29}, 460--493 (2015).

\bibitem{GL}
G.B. Giacomin, J. L. Lebowitz, {\em Phase segregation dynamics in particle systems
with long range interactions. I. Macroscopic limits}, Journal of Statistical Physics  {\bf 87}, 37--61 (1997).

\bibitem{kuramoto} Y. Kuramoto, {\em Chemical oscillations, waves and turbulence}, Springer Series in Synergetics, Springer - Verlag Heidelberg (1984).

\bibitem{LOP} J. L. Lebowitz, E. Orlandi, E. Presutti, {\em A particle model for spinodal decomposition}, Journal of  Statistical Physics {\bf 63}, 933--974 (1991).

\bibitem{LP}  J. L. Lebowitz, O. Penrose, {\em Rigorous Treatment of
the Van Der Waals-Maxwell Theory of the Liquid-Vapor Transition},
Journal of Mathematical Physics {\bf 7}, 98--113 (1966).

\bibitem{presutti} E. Presutti, {\em Scaling limits in statistical mechanics
and microstructures in continuum mechanics} Springer, Theoretical and Mathematical Physics, Springer-Verlag Berlin Heidelberg (2009).

\bibitem{tass} P. A. Tass, {\em Phase resetting in Medicine and Biology: Stochastic Modelling and Data Analysis}, Springer Series in Synergetics, Springer-Verlag Berlin Heidelberg (1999).

\bibitem {wilczek1}  F. Wilczek, {\em Quantum time crystals.} Phys. Rev. Lett. {\bf 109}, 160401 (2012).

\bibitem {wilczek2} A. Shapere, F. Wilczek, Classical time crystals, Phys. Rev. Lett. {\bf 109}, 160402 (2012).

\bibitem{zhang}  J. Zhang, P. W. Hess, A. Kyprianidis, P. Becker, A. Lee, J. Smith, G. Pagano, I.-D. Potirniche, A. C. Potter, A. Vishwanath, N. Y. Yao,  C. Monroe, {\em Observation of a Discrete Time Crystal}. arXiv:1609.08684 (2016).

\end{thebibliography}
\end{document}